\documentclass[12pt]{article}
\usepackage{latexsym}
\usepackage{fancybox,amssymb,color}
\usepackage[dvips]{graphicx}

\let\ssection=\section
\renewcommand{\section}{\setcounter{equation}{0}\ssection}

\newcommand\mathC{\mkern1mu\raise2.2pt\hbox{$\scriptscriptstyle|$}
        {\mkern-7mu\rm C}} % The complex numbers
\newcommand{\mathR}{{\rm I\! R}}         % The real numbers

\newcommand\bi{\begin{itemize}}
\newcommand\ei{\end{itemize}}
\newcommand\be{\begin{equation}}
\newcommand\ee{\end{equation}}

\newcommand{\pl}{\ensuremath{{\partial}}}

\begin{document}
\begin{titlepage}

\begin{center}
{\large\bf Stochastic Einstein Locality Revisited}
\end{center}

\vspace{0.2 truecm}

\begin{center}
        J.~Butterfield \\[10pt] Trinity College, Cambridge CB2 1TQ.\\ Email: jb56@cam.ac.uk
\end{center}

\begin{center}
       Tuesday 14 August 2007
\end{center}

\begin{center}
       Forthcoming in {\em British Journal for the Philosophy of Science} 
\end{center}

\vspace{0.2 truecm}

\begin{abstract}
I discuss  various formulations of stochastic Einstein locality (SEL), which is a version of %%@
the idea of relativistic causality, i.e. the idea that influences propagate at most as fast %%@
as light. SEL is similar to Reichenbach's Principle of the Common Cause (PCC), and Bell's %%@
Local Causality.

My main aim is to discuss formulations of SEL for a fixed background spacetime.  I previously %%@
argued that SEL is violated by the outcome dependence shown by Bell correlations, both in %%@
quantum mechanics and in quantum field theory. Here I  re-assess those verdicts in the light %%@
of some recent literature which argues that outcome dependence does not violate the PCC. I %%@
argue that the verdicts about SEL still stand.

Finally, I briefly discuss how to formulate relativistic causality if there is no fixed %%@
background spacetime.
\end{abstract}

\end{titlepage}

\tableofcontents

\section{Introduction}\label{Intr}
This paper addresses what the the EPR-Bell correlations imply about relativistic causality, %%@
i.e. the requirement that causal influences (signals) can travel at most as fast as light, %%@
within quantum theories. The overall message is that the situation is subtle: various precise %%@
formulations of relativistic causality are satisfied, and others violated, and (more %%@
interestingly) that there remain plenty of open questions.\footnote{Another paper %%@
(Butterfield 2007) focusses instead on less well-known ``single particle'' violations of %%@
relativistic causality. It also locates the discussion within the general philosophy of %%@
causation, especially the causal anti-fundamentalism of Norton (2003, 2006).}

Until the last Section of the paper, I  discuss stochastic relativistic causality for a fixed %%@
relativistic spacetime, in which probabilistic events occur. The spacetime is fixed, in the %%@
sense that these events' outcomes have no influence on the structure of the spacetime. In %%@
this context, the literature considers many different formulations of stochastic relativistic %%@
causality; so I shall have to be very selective. I shall focus on some formulations of %%@
`stochastic Einstein locality' (SEL): an idea which was first proposed by Hellman (1982) for %%@
the special case of Minkowski spacetime as the fixed background. There are three reasons for %%@
this focus.

\indent (i): SEL is a natural expression of the rough idea that causal influences cannot %%@
travel faster than light. Indeed, it is so natural that I think it {\em ought} to be as %%@
familiar to philosophers of probability and causation as is Reichenbach's Principle of the %%@
Common Cause (PCC). We will see, already in Section \ref{sel},  similarities between some %%@
formulations of SEL and PCC; and between SEL and Bell's condition, familiar to physicists, %%@
which he called `local causality'.

\indent (ii): Assessing how both SEL and PCC fare in the Bell experiment leads to some %%@
surprises for the folklore that the Bell experiment refutes PCC. There are two points here. %%@
First: broadly speaking, the arguments for the folklore can be adapted so as to provide %%@
arguments that some natural formulations of SEL are violated by the experiment. So if we want %%@
to express precisely what is ``spooky'' about Bell correlations, the violation of SEL is a %%@
good candidate. Second: recent work by what I shall call `the Budapest school' has objected %%@
to the arguments for the folklore, on the grounds that the arguments assume different %%@
correlations should have the same common cause (now often called a `common common cause'). %%@
The Budapest school backs up this objection with theorems to the effect that  natural %%@
formulations of PCC are {\em obeyed} by the Bell correlations; (they have such theorems for %%@
both elementary quantum mechanics, and quantum field theory). So someone interested in SEL %%@
needs to address the question how this objection, and these theorems, bear on SEL.

\indent (iii): The third reason is personal. In previous work (1989, 1992, 1994, 1996), I %%@
endorsed the folklore that Bell correlations refute PCC, and also argued that they refute %%@
SEL. (Fortunately, I will need very little repetition from that earlier work; nor need I %%@
presuppose it.) So I have a responsibility to consider the Budapest school's objection and %%@
theorems.

There will be two main morals. The first is that there are various formulations of %%@
relativistic causality, indeed of SEL. It is of course no surprise that  formulations will %%@
vary if one considers different conceptions of events and causal influences. But  we will %%@
see, more surprisingly, that even within a fixed conception, the intuitive idea of SEL can be %%@
made precise in inequivalent ways. (But  as we would probably expect: the richer the %%@
conception of events, influences etc., the more inequivalent formulations there are.)

In the two main Sections (\ref{sel} and \ref{qmphys}), I shall  consider a  philosophical %%@
conception of events and influences, close to Hellman's original one (1982). Section %%@
\ref{sel} develops three natural formulations of SEL; and some conditions under which they %%@
are equivalent. In Section \ref{qmphys}, I apply these formulations of SEL to the Bell %%@
experiment; and so turn to topics (ii) and (iii) above. I admit that the situation is not %%@
clear-cut: one has to make  some judgments about how to apply the formulations of SEL to the %%@
experiment, and about setting aside various loopholes e.g. about detector efficiencies. But %%@
given such judgments, one can ask whether the Bell correlations (specifically: outcome %%@
dependence) refute SEL, or PCC.

This yields my second, more specific, moral: despite the work of the Budapest school, a good %%@
case can be made that Yes, the correlations do refute PCC and SEL. Indeed, for PCC in quantum %%@
mechanics, the case has already been made. There are two main points here. First: for the %%@
Bell experiment, it {\em is} reasonable to postulate a common common cause---as the arguments %%@
for the folklore blithely did; (so far as I know, Placek first argued for this). Second: %%@
though the Budapest school is right that Bell's theorems traditionally assumed a common %%@
common cause, there are recent theorems (especially by what I shall call `the Bern school') %%@
assuming only separate common causes for the various correlations. So I will report these two %%@
points, and then adapt them to SEL. (In case this moral sounds defensive, let me confess at %%@
the outset that when in previous work I endorsed the folklore, I was blithely unaware of my %%@
interpreting PCC strongly, viz. as requiring a common common cause: (as, I suspect, were some %%@
other folk). So: {\em mea culpa}, and all credit to the Budapest school for emphasising the %%@
issue.)

In the last two Sections, I briefly discuss how SEL fares beyond quantum mechanics: in %%@
algebraic quantum field theory (AQFT: Section \ref{causeaqft}); and in theories with an {\em %%@
unfixed}, i.e. dynamical, spacetime (Section \ref{causet}). AQFT provides a richer and more %%@
precise conception of events, influences and probabilities than the philosophical conception %%@
of Sections \ref{sel} and \ref{qmphys}. As one would expect, this richer conception provides %%@
many more conditions expressing the broad idea of relativistic causality. But SEL and PCC can %%@
each be applied in a clear-cut way to this conception.  The situation is then broadly as it %%@
was in Section \ref{qmphys}. That is:\\
\indent (a) there are inequivalent formulations of SEL and PCC;\\
\indent (b) the Budapest school shows that a natural formulation of PCC is obeyed by the Bell %%@
correlations; but also;\\
\indent (c) I argue that some natural formulations of SEL are violated.\footnote{A %%@
bibliographic note about how Sections \ref{sel} to  \ref{causeaqft} add to and correct my %%@
previous work; more details in the sequel. Section \ref{sel}'s material adds to the %%@
discussion in my (1994, Section 5).  Section \ref{qmphys} endorses some main claims of my %%@
(1989, Section 2f.; 1992, pp. 74-77; 1994, Sections 6, 7) but corrects others, in the light %%@
of the Budapest school's work. As to AQFT, discussion of SEL in this setting was initiated by %%@
R\'{e}dei (1991), and later work includes Muller and Butterfield (1994). Section %%@
\ref{causeaqft} will endorse my (1996)'s claim that a formulation of SEL is violated by AQFT, %%@
and relate this to R\'{e}dei and Summers' (2002) theorem that a formulation of PCC is obeyed %%@
by AQFT.}

Finally, Section \ref{causet} turns briefly to stochastic relativistic causality in dynamical  %%@
spacetimes. This is a much less well-developed field: in particular, we will see, already in %%@
Sections \ref{sel}, \ref{qmphys} and \ref{causeaqft}, ways in which our formulations of SEL %%@
presuppose a fixed spacetime. For reasons of space, I make just two comments by way of %%@
advertising some open problems. I report an experimental test of whether SEL applies to %%@
metric structure, proposed by Kent; and briefly discuss how SEL fails trivially in the causal %%@
set approach to quantum gravity of Sorkin and others.

\section{Formulating Stochastic Einstein Locality}\label{sel}
After some preliminaries about events and regions (Section \ref{prel}), I will present the %%@
idea of SEL (Section \ref{selintro}), give three precise formulations of it (Section %%@
\ref{3formul}) and discuss implications between them (Section \ref{rel}). The discussion will %%@
be informal and the proofs elementary.

\subsection{Events and regions}\label{prel}
Imagine we are given a spacetime $\cal M$ in regions of which stochastic events occur. Since %%@
closed causal curves notoriously impose consistency conditions on initial data assigned to a %%@
spacelike hypersurface, which can be so severe as to veto the variety of outcomes associated %%@
with a stochastic process, it will be best to assume that $\cal M$ has no closed causal %%@
curves. Indeed, in this Section we shall want to appeal to some spacetime  notions, such as %%@
spacelike hypersurfaces, and some properties of causal ``good behaviour'' enjoyed by some %%@
spacetimes, without going into technicalities about these notions and properties. So for %%@
simplicity, I will assume from the outset that $\cal M$ has the strong good-behaviour %%@
property, {\em stable causality}. I shall not need the exact definition of this; (cf. e.g. %%@
Hawking and Ellis 1973, p. 198; Geroch and Horowitz 1979, p. 241; Wald 1984, p. 198). Suffice %%@
it to say that:\\
\indent (i): The idea is that a spacetime is stably causal if not only does it lack closed %%@
timelike curves, but also the spacetime resulting from a slight opening out of the %%@
light-cones does not have any such curves.\\
\indent (ii): Stable causality has various useful consequences. In particular, a spacetime is %%@
stably causal iff it has a ``global time function'', i.e. a smooth function $f: {\cal M} %%@
\rightarrow \mathR$ whose gradient is everywhere timelike.

For later reference, I should also mention a much stronger good-behaviour notion: a spacetime %%@
is {\em globally hyperbolic} if it has a {\em Cauchy surface}. Here, a Cauchy surface %%@
$\Sigma$ is a ``global instantaneous slice'' in that for every point $p \in {\cal M}$: either %%@
$p$ lies in $\Sigma$'s future {\em domain of dependence} $D^+(\Sigma)$, i.e. every %%@
past-inextendible causal curve through $p$ intersects $\Sigma$; or $p$ lies in the past %%@
domain of dependence $D^-(\Sigma)$, in the corresponding sense that every future-inextendible %%@
causal curve through $p$ intersects $\Sigma$. If a spacetime is globally hyperbolic, then it %%@
is also stably causal; and the global time function $f$ can be chosen so that each surface of %%@
constant $f$ is a Cauchy surface; and $\cal M$ then has the topology $\mathR \times \Sigma$, %%@
where $\Sigma$ denotes any Cauchy surface. Thus a globally hyperbolic spacetime can be %%@
foliated by Cauchy surfaces. (Cf. Hawking and Ellis 1973, p. 205-212; Geroch and Horowitz %%@
1979, p. 252-253; Wald 1984, p. 201, 205, 209.)

Since the events $E, F,...$ occurring in regions of $\cal M$ are to be stochastic, and %%@
influenced by prior events, it is natural to envisage a time-dependent probability function %%@
$pr_t$ which assigns probabilities $pr_t(E)$ etc; with the values of the probabilities %%@
reflecting how the events before $t$ happen to have turned out, and so how the probability of %%@
$E$ has waxed and waned.

 Since $\cal M$ is a relativistic spacetime, we will take $t$ as a hypersurface. Again, I %%@
shall not  worry about the various technical meanings of `hypersurface'.  Suffice it to say %%@
that for the most part, $t$ will be a spacelike hypersurface without an ``edge'', stretching %%@
right across $\cal M$ and dividing it in to two disjoint parts, the ``past'' and ``future''. %%@
However:\\
\indent (a): Not all the hypersurfaces considered will be everywhere spacelike. For we will %%@
allow $t$ to include parts of the boundaries of past light-cones.  Such boundaries are always %%@
3-dimensional embedded sub-manifolds of $\cal M$ which are ``well-behaved'' in being  %%@
achronal, i.e. having no pairs of points $p, q$ connected by a timelike curve (Wald 1984, p. %%@
192).\\
\indent (b): And if $t$ is spacelike, it need not be a Cauchy surface. In particular, for %%@
much of the discussion some weaker notion will suffice, e.g. a closed achronal set without %%@
edge (often called a `slice': Wald 1984, p. 200).

 I shall also use {\em history} for the collection or conjunction of all events up to a given %%@
hypersurface (or within a given spacetime region). This usage will be informal, and in %%@
particular is not meant to have any of the technical connotations of the `consistent %%@
histories' programme in quantum theory. So the probability  distribution $pr_t$ is meant to %%@
reflect how history happens to have turned out (and so modified the prospects of future %%@
events), up to the hypersurface $t$.

There are two issues about the association of an event with a spacetime region: the first %%@
philosophical, the second technical. (Section \ref{causeaqft} will also have more to say %%@
about this association in AQFT.) Most philosophers (including Hellman and those listed in %%@
footnote 2) think of an event as a contingent matter of particular fact that is localized in %%@
a region, by being about properties of objects in that region (or about properties of the %%@
region itself) that are {\em intrinsic} to the region. Philosophers dispute how to analyse, %%@
and even how to understand, the intrinsic-extrinsic distinction among properties. But the %%@
intuitive idea is of course that an intrinsic property implies nothing about the environment %%@
of its instance. For example: `he was fatally wounded at noon' does not just attribute the %%@
intrinsic property of being wounded at noon, since `fatally' implies a later death. On the %%@
other hand, `the average electromagnetic energy density in spacetime region $R$ is $a$' no %%@
doubt does attribute an intrinsic  property to $R$.

I shall not need to be precise about `intrinsic'; nor consequently, about how events are %%@
associated with regions. I shall even use $E$ indifferently for an event and for the %%@
spacetime region with which it is associated.\footnote{But note that some authors' notations %%@
do distinguish events and their associated regions: for example, Henson (2005: Section 2.2.1) %%@
who writes dom($E$) for the region of event $E$, which he calls the `least domain of %%@
decidability of $E$'. Henson thinks of the association epistemically. He says dom($E$) is the %%@
unique smallest region such that knowing all its properties enables us to decide whether $E$ %%@
occurred. I would reply that this epistemic gloss obviously does not avoid the need to %%@
restrict the properties considered to be intrinsic; and that with this restriction, the %%@
epistemic gloss is unnecessary. But this disagreement does not affect the plausibility of %%@
Henson's axioms, which describe how the map dom from events to spacetime regions interacts %%@
with the Boolean operations on events and regions: nor therefore, the results he deduces, %%@
which concern his formulations of PCC.} So to sum up so far: the course of events, or %%@
history, up to a hypersurface $t$ gives a probability $pr_t(E)$ for an event $E$ to occur in %%@
(report some intrinsic properties of) a certain region future to $t$. So intuitively, $E$ is %%@
also a set or union or disjunction of other ways in which history could turn out in its %%@
region, or indeed in the rest of the future of $t$.  

The technical issue is whether we should require an event's spacetime region to have certain %%@
properties, e.g. being open, or bounded, or convex. In fact, Sections \ref{sel} and %%@
\ref{qmphys} will not need any such requirements; but in Section \ref{causeaqft}, AQFT will %%@
work with open bounded regions. But with the exception of one place in Section %%@
\ref{D2impliesD1}'s proof,  we can for the sake of  definiteness, assume throughout Sections %%@
\ref{sel} and \ref{qmphys} that events' regions are bounded and convex. (This simplifies %%@
drawing the diagrams corresponding to the different formulations of SEL.)

A related issue is whether we should define the past light-cone of the region in which an %%@
event $E$ occurs---which will be important for SEL---in terms of the chronological or the %%@
causal past. In fact, the convention is to define these in  slightly non-parallel ways, as %%@
follows; (e.g. Hawking and Ellis 1973, p. 182; Geroch and Horowitz 1979, p. 232; Wald 1984, %%@
p. 190; of course, corresponding remarks hold for futures). The chronological past $I^-(p)$ %%@
of a point $p \in {\cal M}$ is the set of points connectible to $p$ by a  past-directed %%@
timelike curve; $I^-(p)$ is open, but since the curve must be non-zero, in general $p \notin %%@
I^-(p)$. For a set $E \subset {\cal M}$, we define  $I^-(E) := \cup_{p \in E} \; I^-(\{p\})$, %%@
so that $I^-(E)$ is always open. On the other hand, the causal past $J^-(E)$ of a set $E$ %%@
(where maybe $E = \{ p \}$) is defined so as to include $E$: viz. as the union of $E$ with  %%@
the set of points connectible to a point of $E$ by a past-directed causal curve; ($J^-(E)$ %%@
need not be closed). Broadly speaking, it is usually easier to work with chronological pasts %%@
(as Geroch and Horowitz remark). But the results in Section \ref{selintro} et seq. are more %%@
easily obtained if any set $E$ is contained in its past light-cone: which with the %%@
conventional definitions, suggests we should use $J^-(E)$. On the other hand, if $E$ is open %%@
then $E \subset I^-(E)$. So in effect, we have a choice: either we use causal pasts, or we %%@
require $E$ to be open and use chronological pasts. But again, I shall not need to be %%@
precise, and so need not choose. To signal this lack of commitment, I shall adopt the %%@
idiosyncratic notation,  $C^-(E)$, for the past light-cone of $E$ (or of the region in which %%@
an event $E$ occurs). But if one wanted to be definite, one could adopt either of the above %%@
choices.\\
%THINK AGAIN Do I need the region of $E$ to be open and bounded, modulo proof %that SELD2 %%@
implies SELD1? and even also convex? or that the tangent plane to
% the boundary has no timelike directions except in a ``thin equator''?

\subsection{The idea of SEL}\label{selintro}
The idea of relativistic causality can now be expressed as
`Stochastic Einstein Locality' (SEL):
For an event $E$ occurring in the spacetime  $\cal M$,
the probability at an earlier time (hypersurface) $t$ that
$E$ occurs, $pr_t (E)$, should be determined by history (i.e. the events that occurred)
within that part of the past light-cone of $E$ that lies before
$t$; i.e. by history within $C^-(E) \cap C^-(t)$. Here we envisage that $t$:\\
\indent (i) is spacelike, if not everywhere then at least within $C^-(E)$; and \\
\indent (ii) cuts $C^-(E)$ into two disjoint parts, the ``summit'' $C^+(t) \cap C^-(E)$ and %%@
the ``base'' $C^-(t) \cap C^-(E)$.\\
I shall say that a hypersurface $t$ satisfying (i) and (ii) {\em divides} $C^-(E)$.

 So our present aim is to make more precise this idea of determination by the history within %%@
the past light-cone; and in  doing this, we shall see some inequivalent formulations of the %%@
idea. (But we shall maintain the idea that $t$, even if not everywhere spacelike, divides %%@
$C^-(E)$.)

To talk about the various possible total courses of events, it is natural to use %%@
philosophers' jargon of `possible worlds'. A possible world corresponds in physicists' jargon %%@
to a  `solution (throughout all time)' or `dynamically possible total history of the system'; %%@
and in the jargon of stochastic processes, to a `trajectory' or `realization'. I shall write %%@
$w$ for a possible world, and $\cal W$ for the set of possible worlds envisaged.

I should  note here two differences of notation from stochastic process theory. (1): That %%@
theory  usually takes the set of realizations (`worlds') as the basic sample space, with an %%@
event, i.e. a set of realizations, being given by a time-indexed random variable taking a %%@
certain value. This means that although I have written $pr_t (E)$ where it is understood that %%@
$E$ might specify the value, say $q$, of physical quantity $Q$, in stochastic process theory %%@
one writes instead something like $pr(Q_t = q) \equiv pr( \{ {\rm{realizations}} \; w: \; %%@
Q_t(w) = q \})$.\\
\indent (2): Stochastic processes are often assumed to be Markovian. I will not need the %%@
definition of this; but it prompts one to use a notation like my $pr_t$ to represent the %%@
probability prescribed, not by (i) how all of history up until the hypersurface $t$ happens %%@
to turn out, but by (ii) the instantaneous physical state at $t$ itself. So I stress that my %%@
notation $pr_t$ and similar notations below {\em does} mean (i): I shall not need the Markov %%@
property.

It is also natural to use the fact that $\cal M$ is fixed so as to identify times $t$ between %%@
worlds $w \in {\cal W}$, yielding a doubly-indexed probability function $pr_{t,w}$. Here, %%@
identifying times between two worlds does not mean the very suspect idea of a ``meta-time'' %%@
somehow external to both worlds; nor the technical idea from topology of identifying parts of %%@
two spaces to define a single third space. It is to be understood just as the two worlds %%@
matching on all their history up to two hypersurfaces each in its world: given such matching, %%@
the hypersurfaces are then identified in the sense of being both labelled $t$. And here for %%@
two regions in two possible worlds to `match' means that they are isomorphic with respect to %%@
all properties and relations intrinsic to the regions. In the jargon of a physical theory: %%@
there is a smooth bijection between the regions' points (and so between their sub-regions) %%@
that is an isomorphism of all the fields, and whatever other physical quantities, the theory %%@
discusses.\footnote{Two minor points, one specific and one general. (1): We shall see that  %%@
matching in just a past light-cone, rather than in the entire past of a spacelike %%@
hypersurface, is enough for some formulations of SEL. (2): Note that the idea of two worlds %%@
matching up to a time but not thereafter is very different from the idea of a single world %%@
branching or splitting. Though the latter idea has been developed, and used to analyse %%@
quantum nonlocality, by the `Pittsburgh-Krakow school' (cf. footnote 16 for references), I %%@
will not need it in this paper. For cautions about making sense of branching spacetime, cf. %%@
Earman (2006).}

Philosophers will recognize the apparatus I have invoked---events associated with regions, %%@
possible  worlds which can match on regions, and time-dependent objective probability %%@
functions---as reminiscent of David Lewis' metaphysical system. (Cf. especially his (1980, %%@
1986): Lewis calls such probabilities `chances'.) On the other hand, Hellman (1982) does not %%@
use worlds and chances. He talks of formalized physical theories with vocabulary for %%@
probability, and their models. So I should here admit that indeed, philosophical differences %%@
turn on the choices between these frameworks. But fortunately, the differences will not %%@
affect anything in this paper; in particular, nothing in what follows needs any of Lewis' %%@
contentious doctrines about worlds or probabilities.

\subsection{Three formulations of SEL}\label{3formul}
\subsubsection{The formulations}\label{formul}
With these preliminaries, it is easy to write down three different formulations of SEL. The %%@
first concerns the probability of a single event $E$. Both the second and third are %%@
statements that $E$ is stochastically independent of certain other events $F$. Each %%@
formulation will lead on to the next.\footnote{Besides, other authors have other %%@
formulations. My first and third formulations (essentially as in my (1994, Section 5.2)) are %%@
analogous to  Hellman's (1982) two formulations; though he then adds provisos to ensure that %%@
SEL is {\em obeyed} in the Bell experiment; cf.  Section \ref{3302}. My second formulation %%@
below seems to be new. On the other hand, I will not here pursue my (1994, Section 5.2)'s %%@
formulation using counterfactuals with probabilistic consequents.}

First, it is natural to formulate SEL's idea that a probability is {\em determined} by %%@
history within a past light-cone, as follows: for two worlds $w, w' \in {\cal W}$ that match %%@
in their history in $C^-(E) \cap C^-(t)$, the probabilities $pr_{t,w}(E)$ and $pr_{t,w'}(E)$ %%@
are equal. This gives our first formulation of SEL, which I will call `SELS'. The second `S' %%@
stands for `single', since we consider a single event $E$. (For reasons about quantum field %%@
theory, given in Section \ref{causeaqft}, this second `S' also stands for `satisfied'.) Cf. %%@
Figure 1.

\begin{figure}[!htbp]
\centering \fbox{\includegraphics[scale=0.25]{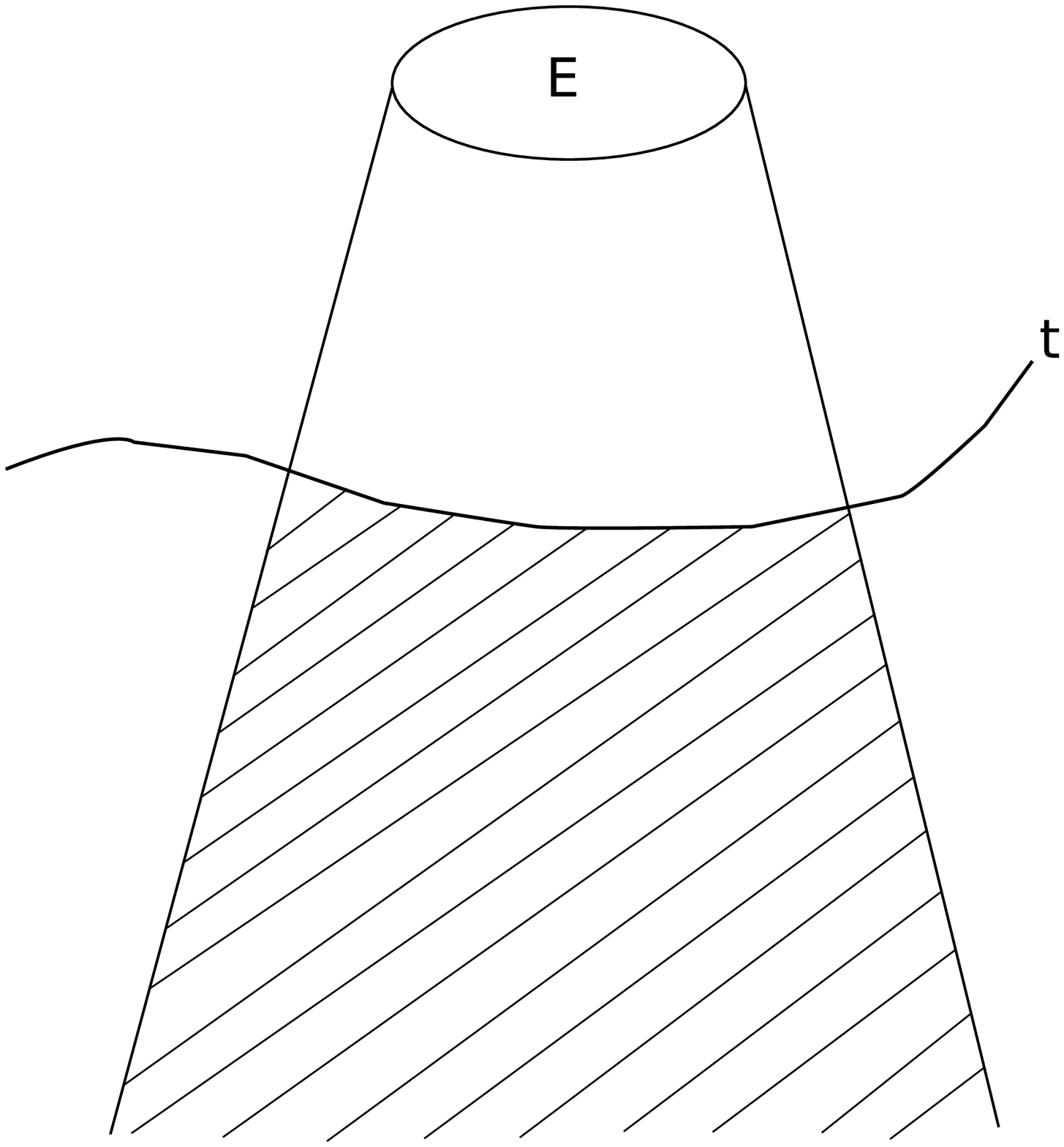}}
\caption{SELS}
\end{figure}

\begin{quote}
{\bf SELS}:
Let worlds $w, w' \in {\cal W}$ match in their history in
 $C^-(E) \cap C^-(t)$. Then they match in their probability at $t$ of $E$:
 \be
pr_{t,w} (E) = pr_{t,w'} (E).
\label{sels}
\ee
Here, as discussed in Section \ref{selintro}: we envisage that the matching histories in $w, %%@
w'$ justify the identification of the two hypersurfaces labelled $t$, and that in both worlds %%@
$t$ divides $C^-(E)$.
\end{quote}

But there is another equally natural approach to formulating SEL. Instead of saying that the %%@
unconditional probability of a  single event $E$ is determined by the ``truncated cone'' of %%@
history in $C^-(E) \cap C^-(t)$, we can instead say that the probability of $E$ is unaltered %%@
by conditionalizing on an event $F$ that occurs in the Elsewhere, i.e. spacelike to $E$. %%@
Here, philosophers will recognize that we connect with Reichenbach's Principle of the Common %%@
Cause (PCC), and its legacy; but I will postpone the comparison with PCC until Section %%@
\ref{comp}.

But we need to be careful about exactly {\em which} events, the given event $E$ is to be %%@
stochastically independent of. For we are considering a probability function that is %%@
time-dependent, and even world-dependent: $pr_t$ or $pr_{t,w}$. So consider an event $F$ that %%@
is future to $t$ and spacelike to $E$'s region, and yet close enough to $E$ that the %%@
intersection of the past light-cones $C^-(E) \cap C^-(F)$ has a part (including its %%@
``summit'') {\em future} to $t$: i.e. $C^-(E) \cap C^-(F) \cap C^+(t) \neq \emptyset$. Such %%@
an event $F$ may well convey information about how history happens to turn out in the future %%@
of $t$, viz. just before the ``summit'' of $C^-(E) \cap C^-(F)$. So $F$ may convey %%@
information about the prospects for whether $E$ occurs---so that $E$ is not stochastically %%@
independent of $F$, according to $pr_t$ (or $pr_{t,w}$). In short: conditioning on such an %%@
$F$ gives information about those of $E$'s antecedents that lie in the future of $t$.

There are two natural solutions to this problem. They will yield our second and third %%@
formulations of SEL.

The first solution  requires that $F$ be not only future to $t$ and spacelike to $E$, but %%@
also far enough away (spatially) from $E$ that $C^-(E) \cap C^-(F) \cap C^+(t) = \emptyset$. %%@
With this requirement, any common causes of $E$ and $F$, whose influence on $E$ or $F$ %%@
propagates at most as fast as light, must lie in the past $C^-(t)$ of $t$. (This follows, %%@
whatever exactly we decide to mean by `common causes', provided they lie in $C^-(E) \cap %%@
C^-(F)$: cf. the second requirement (ii) in the assumption that $t$ divides $C^-(E)$.) So the %%@
idea is that all information about such common causes is already incorporated in $pr_t$; and %%@
so adding that $F$ in fact occurs gives no new information about $E$.
Besides, by making our formulation of SEL consider all hypersurfaces $t$ to the past of $E$, %%@
we can argue that even events $F$, spacelike to $E$  but spatially very close to $E$, will be %%@
required to be stochastically independent of $E$, according to the probability functions %%@
$pr_t$ (or $pr_{t,w}$) for sufficiently late $t$ (i.e. for $t$ so late that $C^-(E) \cap %%@
C^-(F) \cap C^+(t) = \emptyset$).

Thus we get another formulation of SEL. I will call it `SELD1', where the `D' stands for %%@
`double', since we consider two events $E$ and $F$. (For reasons about quantum field theory, %%@
given in Section \ref{causeaqft}, the `D' also stands for `denied'.) Cf. Figure 2.

\begin{figure}[!htbp]
\centering \fbox{\includegraphics[scale=0.25]{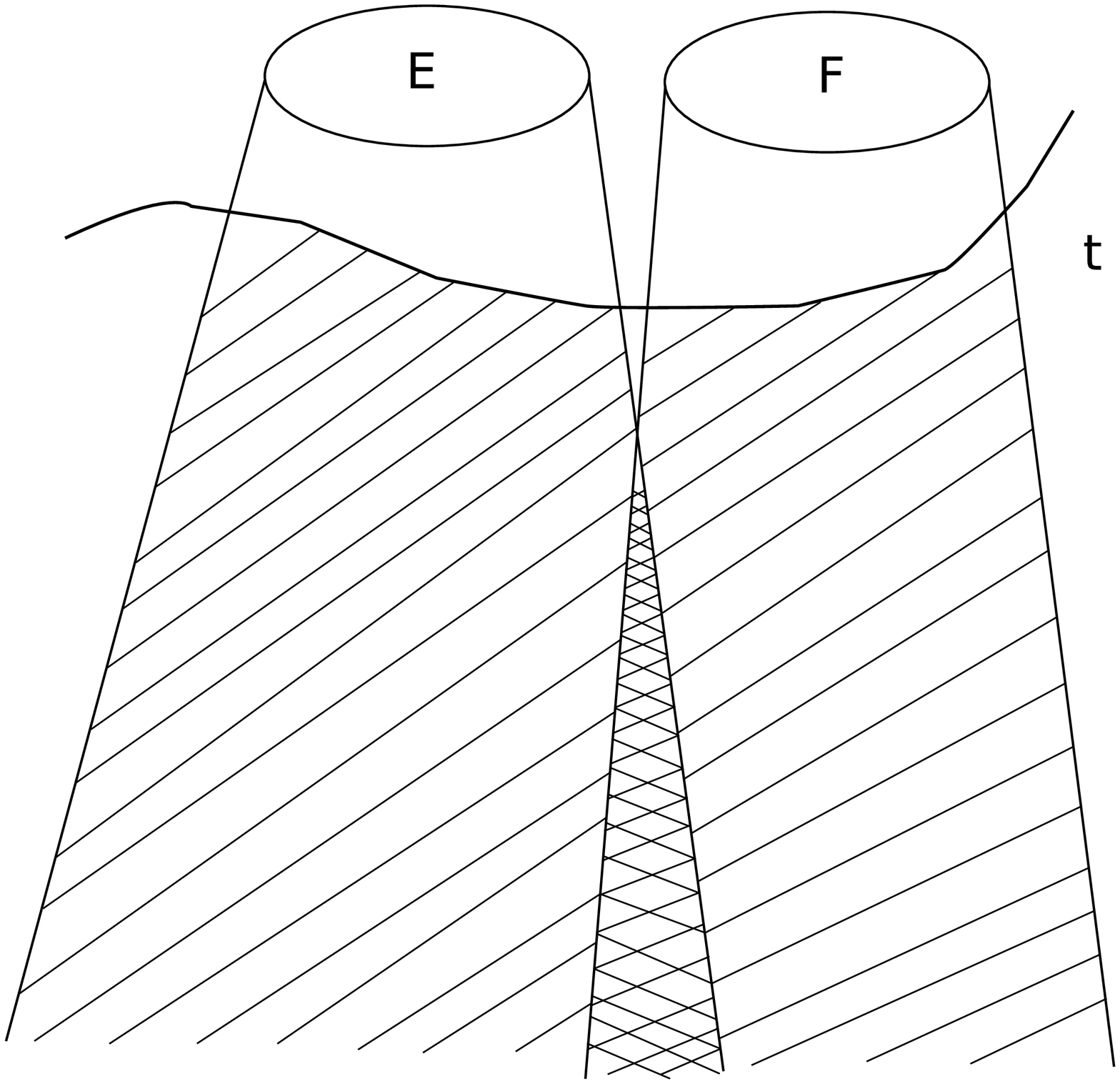}}
\caption{SELD1}
\end{figure}

\begin{quote}
{\bf SELD1}:
For any world $w$, for any hypersurface $t$ earlier than the event $E$ and dividing $C^-(E)$, %%@
and any event $F$ future to $t$ and spacelike to $E$ and such that $C^-(E) \cap C^-(F) \cap %%@
C^+(t) = \emptyset$:
\be
pr_{t,w} (E \& F) = pr_{t,w} (E).pr_{t,w} (F) \; .\footnote{I write eq. \ref{seld1} rather %%@
than $pr_{t,w} (E/F) = pr_{t,w}(E)$, so as to avoid niggling provisos about non-zero %%@
probabilities. I also adopt this tactic in what follows. But anyway such provisos can be %%@
largely avoided in measure theory by using the idea of conditional expectation (e.g. Loeve %%@
1963, p. 341).}
\label{seld1}
\ee
\end{quote}
Remark: One might strengthen SELD1 by dropping the restriction that $F$ be future to $t$, %%@
i.e. by letting $F$ either be in $C^-(t)$ or ``lie across'' $t$. But as will be clear from %%@
our discussion of SELD2, this strengthening is not needed for our purposes.

The idea of the second solution is to require that $F$ be not only spacelike to $E$, but in %%@
the past of $t$, i.e. $F$ lies in the difference $C^-(t) - C^-(E)$. This idea raises three %%@
issues: which I will consider in increasing order of significance for our discussion. Only %%@
the third issue is in any sense a ``problem'': it will lead directly to SELD2. It also %%@
introduces the important  idea that objective probability evolves over time by %%@
conditionalization on how history happens to turn out.

\indent (i): This requirement will exclude any $F$ for which $C^-(E) \cap C^-(F) \cap C^+(t) %%@
\neq \emptyset$. This is so even if, as allowed in Section \ref{prel} , $t$ is not everywhere %%@
spacelike. For the stable causality condition implies that for any region $X$: if $X \subset %%@
C^-(t)$ then $X \cap C^+(t) = \emptyset$; one then applies this to $X := C^-(E) \cap C^-(F)$.

\indent (ii): Recall that SELD1 endeavoured to cover all appropriate events $F$ by %%@
considering all  hypersurfaces $t$ to the past of $E$, even ones just before $E$. So also %%@
here: by considering all such hypersurfaces, we will endeavour to ensure that any $F$ that is %%@
spacelike to $E$ will be in the past of one such hypersurface. After all, such hypersurfaces %%@
can ``tilt up'' towards the future, outside $C^-(E)$, so as to include most of the Elsewhere %%@
of $E$. So any event $F$ which we intuitively want to claim to be stochastically independent %%@
of $E$ {\em can} get included in our formulation of SEL.

\indent (iii): Requiring $F$ to be in the past of $t$, and then considering the conditional %%@
probability function $pr_t( \; /F)$ (or $pr_{t,w}( \; /F)$), raises the question what should %%@
be the objective probability of a past event that actually occurs. The usual answer, in both %%@
the philosophical and technical literature, is that this probability is unity.  This implies %%@
that conditioning $pr_{t}( \; )$ on $F$ makes no difference: that is: $pr_{t}( \; ) = pr_{t}( %%@
\; /F)$. So we cannot express our intuitive idea that $E$ is stochastically independent of %%@
$F$ because $F$ is spacelike to $E$, by using $pr_t$.

 At this stage, the obvious tactic is to use a probability function determined, not by all of %%@
history up to $t$, but by the history lying both in the past of $t$ {\em and} within %%@
$C^-(E)$. So the idea is now that within a world $w$, the history up to $t$ and within %%@
$C^-(E)$ prescribes a probability function, according to which $E$ is stochastically %%@
independent of any possible event $F$ outside $C^-(E)$ but earlier than $t$. Writing $H$ for %%@
this history, and $pr_{H,w}$ for this probability function, we get another formulation (with %%@
the `D2' indicating that this is our second double-event formulation): cf. Figure 3:

\begin{figure}[!hbtp]
\centering \fbox{\includegraphics[scale=0.25]{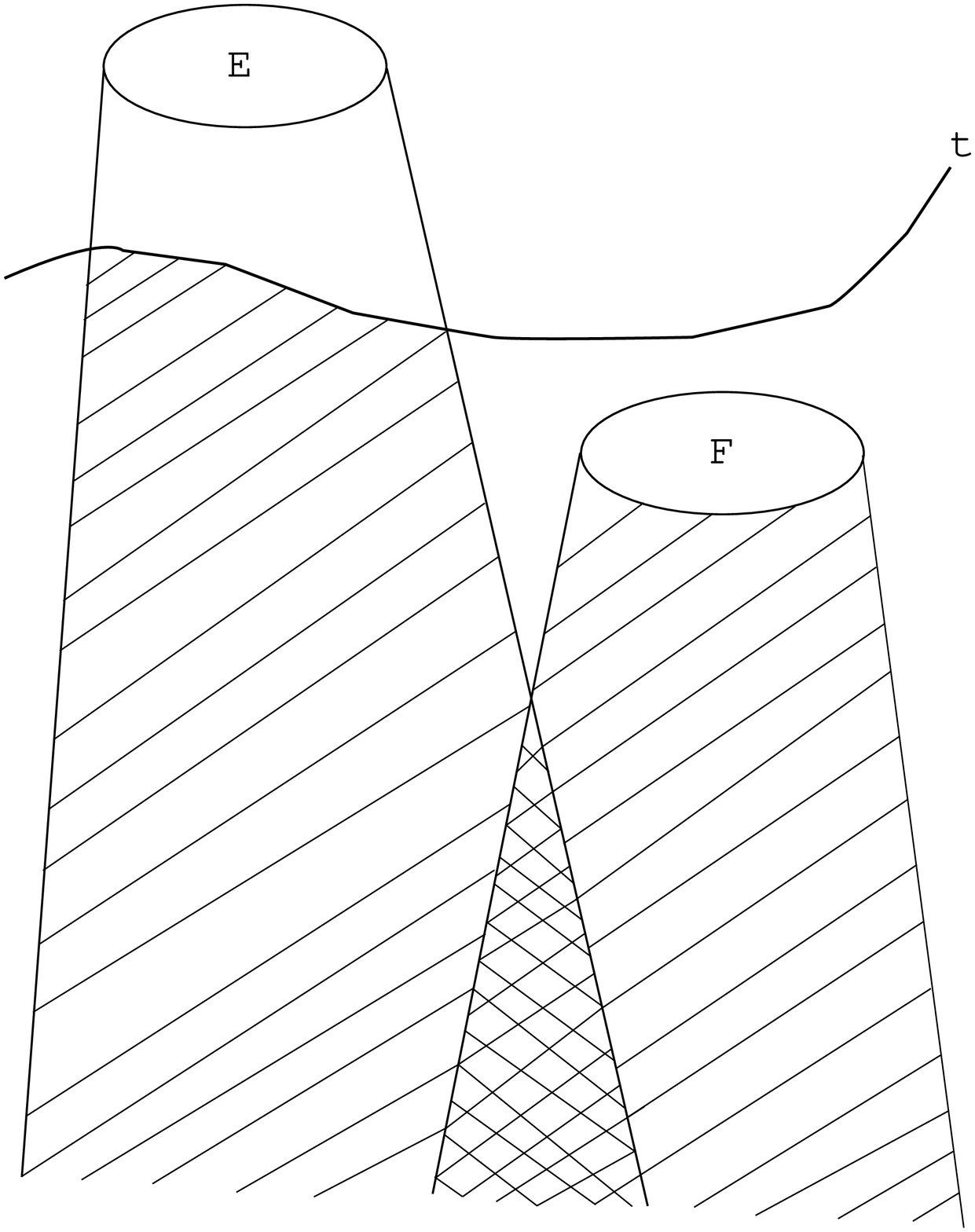}}
\caption{SELD2}
\end{figure}

\begin{quote}
{\bf SELD2}:
For any world $w$, for any hypersurface $t$ earlier than the event $E$ and dividing $C^-(E)$, %%@
and any possible  event $F$ in the difference $C^-(t) - C^-(E)$:
\be
pr_{H,w} (E \& F) = pr_{H,w} (E).pr_{H,w} (F);
\label{seld2}
\ee
where $H$ is $w$'s history in the intersection $C^-(t) \cap C^-(E)$.
\end{quote}
Note that the surface that forms the boundary of the intersection $C^-(t) \cap C^-(E)$ %%@
consists partly of (part of) the boundary of a past light-cone, viz. $C^-(E)$. (The %%@
intersection is like a ``Table Mountain'', whose slopes are part of the boundary of %%@
$C^-(E)$). Since we have written $H$ for the history in this intersection, it is natural to %%@
write $\partial H$ for the boundary, and so to write eq. \ref{seld2} with the surface %%@
$\partial H$ as a subscript, on analogy with SELD1 and eq. \ref{seld1}:
\be
pr_{{\pl H},w} (E \& F) = pr_{{\pl H},w} (E).pr_{{\pl H},w} (F);
\label{seld2dH}
\ee

Finally, there is an important generalization of the idea that the objective probability of a %%@
past event that actually occurs is unity: a generalization which is also usually accepted in %%@
both the philosophical and technical literature. Namely: objective probability evolves over %%@
time by conditionalization on the intervening history. That is to say, in our notation: if %%@
the hypersurface $t$ is later than hypersurface $t'$, and $H^*$ is the conjunction of all %%@
events that occur between $t'$ and $t$, then
\be
pr_{t}( \; ) = pr_{t'}( \; /H^* ) \;.
\label{evolvecond}
\ee
Cf. for example, Lewis (1980, p.101). Of course, any realistic theory is likely to allow %%@
continuously many different possible histories between $t$ and $t'$, so that any one of them %%@
is liable to have probability zero. In that case, eq. \ref{evolvecond} can be made sense of %%@
using conditional expectation; cf. footnote 6.

\subsubsection{Comparisons}\label{comp}
So much by way of stating three formulations of SEL. Before considering their relations %%@
(Section \ref{rel}), I briefly comment on the similarity of the two SELD formulations to (i) %%@
Reichenbach's PCC and (ii) Bell's condition of `local causality'. The main point is of course %%@
that the shared broad idea is so natural that it is no surprise that it is formulated %%@
independently by authors in different literatures: presumably, Bell had never heard of %%@
Reichenbach.\footnote{Similarly, it would be worthwhile to compare our SEL with formulations %%@
of the same broad idea by other authors : for example, Penrose and Percival's principle of %%@
conditional independence (unearthed and endorsed by Uffink 1999) and Henson's (2005) %%@
screening-off conditions. Worthwhile: but we have jobs enough for this paper.}

{\em About PCC}:--- In his PCC, Reichenbach's idea (1956, Section 19) was that if:\\
\indent (i): events $E$ and $F$ are correlated according to some objective probability %%@
function $pr$, $pr(E \& F) \neq pr(E) \cdot pr(F)$; while \\
\indent (ii): there is no direct causal relation between $E$ and $F$:\\
then $E$ and $F$ must be joint effects of a common cause. Reichenbach took this to mean that %%@
there must be a third event $C$ in the ``common past'' of $E$ and $F$ that ``screens them %%@
off'', in the sense that their correlation disappears once we conditionalize on $C$:
\be
pr(E \& F / C) = pr(E / C) \cdot pr(F / C) \; .\footnote{More precisely, Reichenbach   %%@
required, in addition to eq. \ref{screen0}, that: (i) $\neg C$ also screen off $E$ and $F$; %%@
and (ii) $pr(E / C) > pr(E)$ and $pr(F / C) > pr(F)$. From these assumptions, he proves that %%@
$pr(E \& F) > pr(E) \cdot pr(F)$---thus making good the claim that $C$ explains the $E-F$ %%@
correlation. But we can ignore (ii), and treat positive and negative correlations equally; %%@
and we can postpone (i) for a while, viz. until Section \ref{resus} when we will take it in %%@
our stride in discussing common cause variables. Rest assured: both these moves are endemic %%@
in the PCC literature; for example, they are made by Uffink (1999), Henson (2005) and the %%@
Budapest school discussed in Section \ref{B}.}
\label{screen0}
\ee
The idea of eq. \ref{screen0} can be put in vivid epistemic terms, if we imagine an agent %%@
whose initial credence is given by $pr$ and who conditionalizes on what they learn. Namely, %%@
the idea is: since by (ii) $F$ does not causally influence $E$, knowledge that $F$ occurs %%@
gives no further information about whether $E$ will occur, beyond what is already contained %%@
in the knowledge that $C$ occurred; (similarly, interchanging $E$ and $F$). We saw this same %%@
idea in our discussion leading to the SELD formulations, about SELD needing to exclude (from %%@
the claim of stochastic irrelevance to $E$) events $F$ so close to $E$ that their occurrence %%@
gives information about the prospects for $E$, beyond what is already encoded in $pr_t$.

The main similarities and differences between PCC and SELD are clear. Both state  %%@
screening-off between a spacelike/causally unrelated pair of events. But where PCC says that %%@
some past event screens off,  SELD is logically stronger. It is more specific about the %%@
probability function being time-dependent, and so about which events to exclude from the %%@
claim of stochastic irrelevance; (though this can be done in two ways, yielding SELD1 and %%@
SELD2). It also quantifies  universally over earlier times (hypersurfaces) $t$, and so over %%@
histories $H$. It also differs in two other ways:\\
\indent (a): It proposes as the screener-off the entire history up to $t$ (or for SELD2: up %%@
to $t$ and within $C^-(E)$), not some individual event. We will return to this difference, in %%@
Section \ref{prev} and later.  \\
\indent (b): It places its screener-off in a subscript (cf. eq. \ref{seld1}, eq. \ref{seld2}) %%@
labelling a probability function, rather than behind a conditionalization stroke. Again, we %%@
will return to this difference, especially in Sections \ref{314rel.B}, \ref{Bellreview} and %%@
\ref{two}. For the moment, note only that we can already see how this difference might be %%@
elided: cf. eq. \ref{evolvecond}.

{\em About Bell}:--- Bell's `local causality' (2004, p. 54 (2)) is cast in his preferred %%@
language of `beables'. But once translated into Section \ref{prel}'s language of events and %%@
histories, it is very close to SELD, especially SELD2. No surprise: Hellman (1982, p. 466) %%@
and Butterfield (1994, p. 408) admit the inspiration. Thus local causality says: let $E$ and %%@
$F$ be spacelike events in a stochastic theory, let $N$ be the history in the intersection %%@
$C^-(E) \cap C^-(F)$ of their past light cones, let $F'$ be an event in the remainder of %%@
$C^-(F)$, i.e. $C^-(F) - (C^-(E) \cap C^-(F))$: then
\be
\label{Belllc}
pr(E \& F' / N) = pr(E / N) \cdot pr (F' / N).\footnote{In fact, Bell also conditionalizes on %%@
a specification of some beables in the remainder of $C^-(E)$; but I set this aside, i.e. I %%@
take Bell's special case, the ``empty/null specification''. Incidentally, Bell also %%@
emphasizes (2004, pp. 105-106) how countless everyday macroscopic events support the  broad %%@
common idea of PCC and local causality.}
\ee
This is just like SELD2, except that instead of history $H$ within $C^-(t) \cap C^-(E)$ %%@
defining a probability function $pr_H$ that screens off $E$ and $F$ (eq. \ref{seld2}), now %%@
history $N$ within $C^-(E) \cap C^-(F)$ does so---and does so by conditionalizing on $N$ some %%@
given function $pr$; (which is non-time-indexed, presumably because of implicitly assuming %%@
something like eq. \ref{evolvecond}.)
% examine and think about equivalence by choice of the $t$ sweeping up and down;
% mention all worlds have a given $pr$.

\subsection{Implications between the formulations}\label{rel}
These three formulations, SELS, SELD1 and SELD2, are all natural expressions of the idea of %%@
stochastic Einstein locality. But strictly speaking, they  are inequivalent, since they %%@
discuss events in  ``different'' spacetime regions, and different probability functions. %%@
However, I shall state some conditions under which they can be shown equivalent. Section %%@
\ref{314rel.A} concerns the conditions for SELD1 and SELD2 being equivalent; and  Section %%@
\ref{314rel.B} discusses conditions for the equivalence of SELS and SELD2 (and thereby %%@
SELD1).

\subsubsection{Conditions for the equivalence of SELD1 and SELD2}\label{314rel.A}
I shall show that: (i) SELD1 implies SELD2; and  (ii) under certain conditions, SELD2 implies %%@
SELD1. Both proofs will use the fact that by our allowance in (a) of Section \ref{prel}, the %%@
hypersurfaces in SEL can include parts of the boundaries of past light-cones.

\paragraph{SELD1 implies SELD2}\label{D1impliesD2}
\indent Proof: We are given a world $w$, an event $E$, a hypersurface $t$ earlier than $E$ %%@
and dividing $C^-(E)$, and an event $F$ in the difference $C^-(t) - C^-(E)$, and a %%@
probability function $p_{H,w}$ where $H$ is $w$'s history in the intersection $C^-(t) \cap %%@
C^-(E)$.\\
\indent We note that since $F \subset C^-(t)$, $C^-(F) \cap C^-(E) \subset C^-(t) \cap %%@
C^-(E)$. But this latter is the region associated with the history $H$. So SELD1 applies, but %%@
now taking  as the hypersurface in SELD1, the boundary of $C^-(t) \cap C^-(E)$. Thus SELD1's %%@
eq. \ref{seld1} becomes SELD2's eq. \ref{seld2}, or with the boundary $\pl H$ as label, eq. %%@
\ref{seld2dH}: as desired. QED.

To prove the converse, that SELD2 implies SELD1 (under certain conditions), we first need a %%@
lemma which states a ``concordance'' between SELD1's and SELD2's conditions on spacetime %%@
regions: i.e. on the spatiotemporal, especially causal, relations between $E, F$ and $t$. (So %%@
this concordance is not about their requirements on probability functions.)
\begin{quote}
{\bf Concordance}: For any event $E$, any hypersurface $t$ earlier than $E$ and dividing %%@
$C^-(E)$, and any event $F$ spacelike to $E$:\\
If $C^-(E) \cap C^-(F) \cap C^+(t) = \emptyset$, then there is a hypersurface $t'$ that %%@
coincides with $t$ across all of $C^-(E)$ (and so also divides $C^-(E)$) and is such that $F %%@
\subset C^-(t')$.
\end{quote}
Proof: We use the fact that by (a) in Section \ref{prel}, our hypersurfaces can include parts %%@
of the boundaries of past light-cones. Thus we construct $t'$ by defining it to be the %%@
boundary of the region $[(C^-(E) \cap C^-(t)) \cup C^-(F)]$. (So this region is like a %%@
``table mountain'' $(C^-(E) \cap C^-(t))$ beside a ``summit'' $C^-(F)$. Note that because %%@
$C^-(E) \cap C^-(F) \cap C^+(t) = \emptyset$, $F$ cannot be so close to $E$ that the %%@
``table-top'' becomes a ``mere ledge'' on the slope of the mountain $C^-(F)$.) Or %%@
alternatively we can define $t'$ to consist of the boundary of $C^-(t) \cup C^-(F)$: this %%@
will make $t'$ coincide with $t$ everywhere, except at the ``summit'' of $C^-(F)$, where $t'$ %%@
will be the surface of the summit. QED.

(Though it is not needed for what follows, we remark that one can prove the converse of %%@
Concordance, by showing that if $t$ and $t'$ coincide in $C^-(E)$ and both divide $C^-(E)$, %%@
then $C^+(t') \cap C^-(E) = C^+(t) \cap C^-(E)$. So if the intersection of $C^-(F)$ with the %%@
left hand side is empty, i.e. $C^-(F) \cap C^+(t') \cap C^-(E) = \emptyset$, then the %%@
intersection of $C^-(F)$ with the right hand side: yielding the converse.)

\paragraph{SELD2 implies SELD1, given two assumptions}\label{D2impliesD1}
It will be clearest to introduce the assumptions we need, by embarking on a proof of the %%@
implication and  introducing them as the need for them becomes clear.

\indent So we are given a world $w$, an event $E$, a hypersurface $t$ earlier than $E$ and %%@
dividing $C^-(E)$, and an event $F$ future to $t$, spacelike to $E$ and such that  $C^-(E) %%@
\cap C^-(F) \cap C^+(t) = \emptyset$. We seek to show
\be
pr_t(E \& F) = pr_t(E).pr_t(F) \; ;
\label{seekseld1}
\ee
where we have suppressed the world index from eq. \ref{seld1}.

First, we infer from Concordance that  there is a hypersurface $t'$ that coincides with $t$ %%@
across all of $C^-(E)$ and such that $F \subset C^-(t')$. So SELD2 applies, with $H$ being %%@
the world $w$'s history in the intersection $C^-(t) \cap C^-(E) \equiv C^-(t') \cap C^-(E)$. %%@
That is, we have  eq. \ref{seld2}, i.e. again suppressing the world-index:
\be
pr_{H} (E \& F) = pr_{H} (E).pr_{H} (F) \; .
\label{seld2inproof}
\ee
Or in terms of the surface $\pl H$, we have eq. \ref{seld2dH}:
$pr_{{\pl H}} (E \& F) = pr_{{\pl H}} (E).pr_{{\pl H}} (F).$

But to get eq. \ref{seekseld1}, we still need to evolve probabilities forward from the %%@
hypersurface $\pl H$, i.e. the boundary of $C^-(t) \cap C^-(E) \equiv C^-(t') \cap C^-(E)$, %%@
to $t$ itself. That is: we need to show that the stochastic independence in eq. %%@
\ref{seld2inproof} is not lost as one evolves forward from $\pl H$ to $t$.

To show this, we need further assumptions. After all, events in $C^-(F) - C^-(E)$ can %%@
perfectly well influence $F$ subluminally; and for all we have so far said, they might spoil %%@
the stochastic independence in eq. \ref{seld2inproof}.

There are no doubt various assumptions that would secure the implication. I think it most %%@
natural to make the following two assumptions. (For the argument to follow, we will no longer %%@
need to consider the hypersurface $t'$ inferred from Concordance.) The first is entirely %%@
general: it is that probabilities evolve by conditionalization on the intervening history, %%@
i.e. eq. \ref{evolvecond}.

\indent The second assumption is specific to SEL. It is an equation of ratios of %%@
probabilities, viz. eq. \ref{plaus?} below. But it will be clearest to introduce it when we %%@
see its role in the proof. Suffice it to say initially that the equation combines two ideas %%@
about events $E$ and $F$, and a hypersurface $t$, given as in the hypothesis of SELD1:\\
\indent (a): The events in $C^-(t) \cap (C^-(F) - C^-(E))$ contain exactly the same %%@
influences on $F$ (perhaps more neutrally: contain all the information about the prospects %%@
for $F$ happening) as are contained in the events in $C^-(t) - C^-(E)$. The latter region is %%@
larger since for any sets $X, Y, Z: X \cap ( Y - Z) \subset X - Z$. But the events in the %%@
difference, i.e. in $C^-(t) - C^-(E)$ but not $C^-(t) \cap (C^-(F) - C^-(E))$, are spacelike %%@
to $F$. So this idea reflects the prohibition of superluminal influences on $F$.\\
\indent (b): Neither of the regions, $C^-(t) \cap (C^-(F) - C^-(E))$ and $C^-(t) - C^-(E)$,  %%@
contains influences on $E$. This reflects the prohibition of superluminal influences on $E$.

So now let us assume that probability functions evolve by conditionalization on intervening %%@
history in the sense of eq. \ref{evolvecond}. Then if $H^*$ is $w$'s history in the %%@
difference $C^-(t) - C^-(E)$, and (as before) $H$ is the history in the intersection $C^-(t) %%@
\cap C^-(E)$,  we have:
\be
pr_{t,w} ( \; ) = pr_{H,w}( \; / H^*).
\label{evcond}
\ee
We  combine this with the instance of SELD2 that takes $F$ in SELD2 to be the {\em whole} of %%@
the intervening history $H^*$ in the difference $C^-(t) - C^-(E)$. So eq. \ref{seld2} (i.e. %%@
\ref{seld2inproof}) and \ref{evcond} gives
\be
pr_{t,w}(E) = pr_{H,w}( E / H^*) = pr_{H,w}(E) \; .
\label{neat}
\ee

Let us now write $H_E$, rather than our previous $H$, for the history in $C^-(t) \cap %%@
C^-(E)$. Similarly let $H_F$ be the history in $C^-(t) \cap C^-(F)$. (We will not need the %%@
$t$ label to be explicit in such $H_E, H_F$; nor will we need the world-index.) Then eq. %%@
\ref{neat} can be written as:
\be
pr_t(E) = pr_{H_E}(E) \; .
\ee
An exactly similar argument for the event $F$ given in the hypothesis of SELD1 (i.e. future %%@
to $t$, spacelike to $E$, and such that  $C^-(E) \cap C^-(F) \cap C^+(t) = \emptyset$) yields
\be
 pr_t(F) = pr_{H_F}(F) \; .
\ee
So what we seek, eq. \ref{seekseld1}, becomes
\be
pr_t(E \& F) = pr_{H_E}(E).pr_{H_F}(F) \; .
\label{seekseld12}
\ee
And what we know, SELD2, eq. \ref{seld2}, is now
\be
pr_{H_E}(E \& F) = pr_{H_E}(E).pr_{H_E}(F) \; .
\label{seld22}
\ee

Now we  see that we can complete the proof by assuming an equation of ratios of %%@
probabilities; as follows. Since $pr_{H_E}(E)$ is in both eq. \ref{seekseld12} and %%@
\ref{seld22}, eq. \ref{seekseld12} will follow immediately if we assume:
\be
\frac{pr_t(E \& F)}{pr_{H_F}(F)} = \frac{pr_{H_E}(E \& F)}{pr_{H_E}(F)} \; .
\label{plaus?1}
\ee
Equivalently, it suffices to assume:
\be
\frac{pr_t(E \& F)}{pr_{H_E}(E \& F)} = \frac{pr_{H_F}(F)}{pr_{H_E}(F)} \; .
\label{plaus?}
\ee

Eq. \ref{plaus?} is, I submit, a plausible assumption. Though it equates relative %%@
probabilities (i.e. ratios of probabilities), the obvious justification for it lies in %%@
identifying two bodies of information (i.e. sets of events), one associated with $C^-(t) - %%@
C^-(E)$ and the other with $[(C^-(t) \cap C^-(F)) - (C^-(t) \cap C^-(E))] \equiv [C^-(t) \cap %%@
(C^-(F)) - C^-(E))]$. For short, I will temporarily write this latter region as $[F-E]_t$. %%@
Note that this region is ``most'' of $C^-(F) - C^-(E)$: think of how in a two-dimensional %%@
spacetime diagram, $C^-(F) - C^-(E)$ forms an diagonal strip extending infinitely in to the %%@
past; $[F-E]_t$ is this strip apart from its ``summit'' future to $t$. The justification of %%@
eq. \ref{plaus?} now goes as follows.

\indent (i): The information about the prospects for $E \& F$ happening, that is associated %%@
with $C^-(t) - C^-(E)$, is represented by the left hand side of eq. \ref{plaus?}. (For $H_E$ %%@
is the history in $C^-(t) \cap C^-(E)$.) If the left hand side is greater than 1, then that %%@
information promotes $E \& F$  in the sense that the events in $C^-(t) - C^-(E)$ make $E \& %%@
F$ more likely.\\
\indent (ii): Similarly: the information about the prospects for $F$ happening, that is %%@
associated with $[F-E]_t$, is represented by the right hand side of eq. \ref{plaus?}. If the %%@
right hand side is greater than 1, then that information promotes $F$  in the sense that the %%@
events in $[F-E]_t$ make $F$ more likely.

Why should these quantitative measures of promotion (or inhibition) be the same? Surely the %%@
most natural justification is that these two bodies of information (sets of events) {\em are %%@
the same}. That is: the information about the prospects for $E \& F$ happening, that is %%@
associated with $C^-(t) - C^-(E)$, is exactly the information about the prospects for $F$ %%@
happening, that is associated with $[F-E]_t$. As I mentioned in (a) and (b) just above eq. %%@
\ref{evcond}, this identity reflects the prohibition of superluminal influences on $F$, just %%@
as much as on $E$.

 So to sum up the justification of eq. \ref{plaus?} (equivalently: \ref{seekseld1} and %%@
\ref{seekseld12}):--- The prohibition of superluminal influences makes these two bodies of %%@
information the same; and if  they are the same, the prospects for the respective events will %%@
match, giving eq. \ref{plaus?}. For example, if this information is ``positive'', i.e. %%@
promotes the events $E \& F$ and $F$ respectively, then the ratio in eq. \ref{plaus?} will be %%@
greater than one. QED.

\subsubsection{Conditions for the equivalence of SELS and SELD2}\label{314rel.B}
I turn to stating some conditions under which SELS and SELD2 are equivalent.\footnote{What %%@
follows repeats my (1994, pp. 409-410); but is worth repeating, not just for the sake of %%@
completeness but also for some points in Sections \ref{qmphys} and \ref{causeaqft} .} (So %%@
using Section \ref{314rel.A}, we could  deduce conditions under which SELS and SELD1 are %%@
equivalent.)

Suppose we assume that:\\
\indent (i) probabilities evolve by conditionalization eq. \ref{evolvecond};\\
\indent (ii) all the worlds in $\cal W$ have the same initial probability function $pr$; this %%@
will mean that we do not need a world-subscript $w$; (Lewis considers this: 1980, p. 112, %%@
131).\\
It is then easy to show that SELS and SELD2 are equivalent, if we make the technically %%@
simplifying pretence that the worlds of $\cal W$ have:\\
\indent (a) only finitely many possible histories, $G_i$ say, for the region $C^-(t) - %%@
C^-(E)$; and \\
\indent (b) only finitely many possible histories, $H_k$ say, for the region $C^-(t) \cap %%@
C^-(E)$.\\
(To generalize the proof that follows to realistic numbers of histories would require us to %%@
apply conditional expectation to extremely large measure spaces, cf. footnote 6: but I shall %%@
not attempt this.)

Assumptions (i) and (ii) imply that: in SELS, we can write $pr_{t,w}(E)$ as $pr(E / G_i \& %%@
H_k)$, where $w$ has histories $G_i$ and $H_k$ up to $t$; and in SELD2, we can write %%@
$pr_{H,w}(E)$ as $pr(E / H_k)$. The proof then applies the elementary result that for any %%@
probability function $p$, with $\{ Y_i \}$ as a partition of its space and the $p(Y_i)$ %%@
non-zero, and for any event $X$:
\be
[\forall \; i, \; p(X/Y_i) = p(X)] \;\;\; {\rm {iff}} \;\;\;
[\forall \; i \; {\rm {and}} \; j, \; p(X/Y_i) = p(X/Y_j)] \; .
\label{elyequivce}
\ee
To apply eq. \ref{elyequivce}, we note that assumptions (i) and (ii) mean that eq. \ref{sels} %%@
becomes
\be
\forall k,i,i': \;\; pr(E / G_i \& H_k) \; = \; pr(E / G_{i'} \& H_k) \; .
\label{SELSbecome}
\ee
By eq. \ref{elyequivce}, this is so iff
\be
\forall k,i: \;\; pr(E / G_i \& H_k) \; = \; pr(E / H_k) \; .
\label{special}
\ee
This gives eq. \ref{seld2}, i.e. SELD2, for the special case where $F$ is a maximally strong %%@
event, one of the $G_i$. For the general case, we treat $F$ as the exclusive disjunction of %%@
the various total histories $G_i$ that include it (in effect: include it as a conjunct), and %%@
we sum over this limited range of $G_i$. Thus in any world with history $H$ within $C^-(t) %%@
\cap C^-(E)$:
\begin{eqnarray}
pr(E / F \& H) = pr(E \& F / H) \; / \; pr(F/H) \\ \nonumber
 = \left[ \Sigma_i \; pr(E \& G_i / H) \right] \; / \; \left[ \Sigma_i \; pr(G_i / H) \right] %%@
\\ \nonumber
= \left[ \Sigma_i \; pr(E / G_i \& H) \right] \cdot pr(G_i / H) \; / \; \left[ \Sigma_i \; %%@
pr(G_i / H) \right] \\ \nonumber
= pr(E / H) \cdot \left[ \Sigma_i \; pr(G_i / H) \right] \; / \; \left[ \Sigma_i \; pr(G_i / %%@
H) \right] \\ \nonumber
= pr(E / H) \; ;
\label{94proof}
\end{eqnarray}
where the last line applies the special case eq. \ref{special} already obtained. For the %%@
converse entailment, from SELD2 to SELS, we again take the special case of eq. \ref{seld2} %%@
where $F$ is a $G_i$. That is, we take eq. \ref{special}. Then we apply eq. \ref{elyequivce} %%@
to get eq. \ref{SELSbecome}, i.e. SELS.

\section{Relativistic causality in the Bell experiment}\label{qmphys}
So much by way of introducing various formulations of SEL. In this Section and the next, I %%@
will discuss how these formulations fare in quantum physics: first, in the Bell experiment of %%@
elementary quantum mechanics (this Section); and then in quantum field theory, in its %%@
algebraic formulation  (AQFT: Section \ref{causeaqft}). For both quantum mechanics and AQFT, %%@
the gist of the discussion will be that violation of outcome independence (a central %%@
assumption in a proof of a Bell inequality) implies that one or another formulation of SEL is  %%@
violated.

In this Section, I begin by reviewing the Bell experiment, and some ``ancient'' discussions %%@
by me and others of SEL and PCC (Section \ref{sofar}). In Section \ref{B}, I first review the %%@
Budapest school's resuscitation of the PCC, based on the distinction between every %%@
correlation having a common cause, and all correlations having the {\em same} such cause. %%@
Then I endorse the recent arguments of Placek and the Bern school that, after all, outcome %%@
dependence does impugn PCC. Finally in Section \ref{SELBE}, I carry over these arguments to %%@
SEL: after the long Section \ref{B}, that will be easy work. It will also be clear that the %%@
discussion generates some open questions.

\subsection{The background}\label{sofar}
Section \ref{Bellreview} describes local models of the Bell experiment in terms which are %%@
both common, and needed for the reviews of  previous work: both my own in Section \ref{prev}, %%@
and others' in Sections \ref{B}.

\subsubsection{The Bell experiment reviewed}\label{Bellreview}
A stochastic local model of the spin (or polarization) version of the traditional two-wing %%@
Bell experiment postulates a space $\Lambda$ of complete states of the pair of particles.  We %%@
represent the two possible choices of measurement (of spin-component) on the left (L) wing by %%@
$a_1, a_2$; and on the right (R) wing, by $b_1, b_2$. The idea is that $\lambda \in \Lambda$  %%@
encodes all the factors that influence the measurement outcomes that are settled before the %%@
particles  enter the apparatuses, and that are therefore not causally or stochastically %%@
dependent on the measurement choices: (we will later be more precise about this). So a state %%@
(``hidden variable'') $\lambda$ specifies probabilities for outcomes $\pm 1$  of the various %%@
single and joint  measurements:
\be
pr_{\lambda, a_i}(\pm 1) \; , \; pr_{\lambda,  b_j}(\pm 1) \; , \; {\rm{and}} \;\;\; %%@
pr_{\lambda, a_i, b_j}(\pm 1 \& \pm 1) \; ; i,j = 1,2.
\label{atlambda}
\ee
We also represent measurement outcomes by $A_i, B_i, i = 1,2$, where $A_i = \pm 1$ is the %%@
event that measuring $a_i$ has the outcome $\pm 1$.  We  will also use $x$ as a variable over %%@
$a_1, a_2$;  $X$ as a variable over $A_1, A_2$ and their negations (i.e., outcome -1); and %%@
for the right wing, we similarly use $y$ and $Y$.

Observable probabilities are predicted by averaging over $\lambda$. For example, the %%@
observable  left wing single probability for $A_1 = + 1$ is:
\be
pr(A_1 = + 1)  := \int_{\Lambda} \; pr_{\lambda, a_1}(+ 1) \; d \rho \; .
\ee
(Some treatments include apparatus hidden variables, i.e. factors in each apparatus %%@
influencing the outcome, so that $pr_{\lambda, a_1}$ is itself an average over hidden %%@
variables $\lambda_L$ associated with the L-apparatus, using some distribution $\rho_L$ say: %%@
$pr_{\lambda, a_1}(X) = \int pr_{\lambda, a_1, \lambda_L}(X) \; d \rho_L $. A Bell inequality %%@
is still derivable, given that the L and R apparatus hidden variables are suitably %%@
independent.  But I think my arguments would be unaffected by this complication, and I set it %%@
aside.)

Note that eq. \ref{atlambda}'s subscript notation prompts one to think of joint measurements %%@
in terms of four copies of the probability space that is the Cartesian product of $\Lambda$ %%@
and the tiny 4-element space $\{<+1,+1>, ..., <-1,-1> \}$; where each copy is labelled by one %%@
of the four joint choices $<a_1,b_1>, ..., <a_2,b_2>$ of measurement, and the measure on the %%@
probability space is fixed by $\rho$ and the joint probabilities in  eq. \ref{atlambda}. %%@
Similarly, of course for single measurements. In other words: we can think of joint %%@
measurements in terms of four vector-valued random variables $<a_1,b_1>, ..., <a_2,b_2>$ on %%@
the space $\Lambda$, all four random variables having $\{<+1,+1>, ..., <-1,-1> \}$ as %%@
codomain; (and similarly for single measurements, with each of $a_1, a_2, b_1, b_2$ a random %%@
variable having $\{+1, -1 \}$ as codomain).

Since philosophers  tend to think that not every event (or fact or proposition) has a %%@
probability, this ``many spaces approach'' has the advantage that by not representing the %%@
choices $a_1,...,b_2$ as events in a probability space, they do not need to be assigned a %%@
probability.  But it has the disadvantage of making it harder to say that there is no causal %%@
or stochastic dependence between the choices and other events or facts, especially the value %%@
of $\lambda$. This disadvantage can be overcome by using a ``big space'', whose elements are %%@
ordered quintuples $<\lambda, x, y, X, Y >$; with the variables $x, y, X, Y$ understood as %%@
above. We can then say, for example, that each value of $\lambda$ is stochastically %%@
independent of each measurement choice:
\be
pr(\lambda \& x) = pr(\lambda) \cdot pr(x) \;\;\; ; \;\;\; pr(\lambda \& y) = pr(\lambda) %%@
\cdot pr(y) \; .
\label{lambchoice}
\ee
The measure $pr$ here can be defined straightforwardly from $\rho$, the probabilities given %%@
by each $\lambda$ i.e. eq. \ref{atlambda}, and postulated probabilities for the measurement %%@
choices $x, y$. For the moment, I will continue with the many spaces approach, but in %%@
Sections \ref{two} and \ref{sz}, we will return to the big space approach. (For more %%@
discussion of the approaches, cf. Butterfield (1989, p. 118; 1992, Sections 2, 3), Berkovitz %%@
(1998a Section 2.1; 2002, Section 3.3; 2007, Section 2).)

We now assume ``locality'' in two ways. First: the measure $\rho$, by which we average over %%@
$\lambda$ to get observable probabilities, is independent of the measurement choices: we do %%@
not write $\rho_{x,y}$ or, for a single L measurement, $\rho_x$.\footnote{But we will later %%@
discuss  how such dependence is correct, if $\lambda$ represents, not just all causal factors %%@
that are settled before the particles enter the apparatuses, but also the measurement %%@
choices. It can be convenient to adopt this approach: cf. Section \ref{sz}'s discussion of  %%@
Szab\'{o} (2000).}

Second: the joint probabilities prescribed by each value of $\lambda$ are assumed to %%@
factorize into the corresponding single probabilities:
\be
\forall \lambda; \forall x,y; \forall X, Y = \pm 1: \;\;
pr_{\lambda, x,y}(X \& Y) =  pr_{\lambda, x}(X) \cdot pr_{\lambda,  y}(Y) \; .
\label{facby}
\ee
Eq. \ref{facby} is called `factorizability' or `conditional stochastic independence'. We will %%@
of course return to its relation to the idea of $\lambda$ as a common cause of the  outcomes, %%@
and so to the PCC and SEL. For now, we note that it  is the conjunction of two disparate %%@
independence conditions; (for the proof, cf. e.g. Redhead (1987, pp. 99-100), Bub (1997, p. %%@
67)). The first is, roughly speaking, independence from the measurement  choice in the other %%@
wing; called `parameter independence' (where `parameter' means `apparatus-setting'):
\be
\forall \lambda; x,y; X, Y = \pm 1: \;\;
pr_{\lambda, x}(X) = pr_{\lambda, x,y}(X) :=   pr_{\lambda, x,y}(X \& Y) + pr_{\lambda, %%@
x,y}(X \& \neg Y) \; ;
\label{pi}
\ee
and similarly for R-probabilities. The second condition is, roughly, independence from the %%@
outcome obtained in the other wing: `outcome independence':
\be
\forall \lambda, x,y; X, Y = \pm 1: \;\;
pr_{\lambda, x,y}(X \& Y) =  pr_{\lambda, x,y}(X) \cdot pr_{\lambda, x,y}(Y) \; .
\label{oi}
\ee

Bell's theorem states that any stochastic local model in this sense---especially, obeying eq. %%@
\ref{facby}, or equivalently, eq. \ref{pi} and \ref{oi}---is committed to a Bell inequality %%@
governing certain combinations of probabilities (e.g. Redhead (1987, pp. 98-101), Bub (1997, %%@
pp. 56-57), Shimony (2004, Section 2)): which is experimentally violated.  On the other hand, %%@
quantum theory is not thus committed, and many experiments confirm the quantum theoretic %%@
predictions.

So which assumption of the derivation of a Bell inequality, or of its experimental testing, %%@
is the culprit? The usual verdict is: outcome independence; and in this paper I will endorse %%@
this verdict, so as to pursue the consequences for PCC and SEL. Indeed, this verdict is %%@
suggested by quantum theory's obeying parameter independence but not outcome independence. %%@
More precisely: putting a quantum mechanical state for each $\lambda$, and taking the %%@
probabilities at a given $\lambda$, eq. \ref{atlambda}, to be given by the orthodox Born %%@
rule, we infer that:\\
\indent (i): eq. \ref{pi} holds: it is now a statement of the quantum no-signalling theorem, %%@
following from the commutation of the L- and R-quantities; but \\
\indent (ii): eq. \ref{oi} fails, except in some special cases such as the model's quantum %%@
state being a product state. \\
Besides, quantum theoretic descriptions of a much less idealized Bell experiment retain this %%@
contrast between parameter independence and outcome independence: that the first holds, and %%@
the second fails.

But I emphasise that this usual verdict is a matter of judgment; for two reasons. First, %%@
rival interpretations of quantum theory (especially: approaches to the measurement problem) %%@
can motivate different verdicts. The best-known example is the pilot-wave approach. Here it %%@
is natural to take $\lambda$ to include both the quantum state, and the ``hidden variables'', %%@
e.g. point-particle positions in elementary quantum mechanics: in which case, parameter %%@
independence {\em fails} and outcome independence {\em holds}. The quantum no-signalling %%@
theorem is nevertheless recovered at the level of observable probabilities by averaging over %%@
the hidden variables  using the ``quantum equilibrium'' distribution (i.e. the Born rule). %%@
Other examples include dynamical reduction theories and many-worlds interpretations: even %%@
though in some cases of these examples, the one-liner verdict is the same as above---i.e. %%@
outcome independence fails---the interpretation of this failure can be very different. (Cf %%@
e.g.: Butterfield et al. (1993) and Kent (2005) for dynamical reduction theories; and %%@
Bacciagaluppi (2002, Sections 4, 6) and Timpson and Brown (2002, Section 4) for many-worlds %%@
interpretations.) And these different interpretations can have the merit of revealing others' %%@
implicit assumptions. No doubt, the main case of this is how many-worlds interpretations %%@
reveal the implicit assumption I have made, that measurement outcomes are genuine, definite %%@
events.\footnote{But some philosophers have been prompted by desiderata in the philosophy of %%@
causation, rather than by many-worlds interpretations, to articulate, and even argue for, %%@
this assumption; e.g. Butterfield (1994, Section 4), Grasshoff et al (2005, p. 668).}

Second, people can of course differ about which assumptions of Bell's theorem are false,  %%@
without having to motivate their opinion by anything so ambitious as an interpretation of %%@
quantum theory. The two best-known  ways to save factorizability eq. \ref{facby}, are usually %%@
called the `detector efficiency' and `locality' loopholes. I will ignore the former, which is %%@
usually considered almost  closed, and concentrate on the latter.\footnote{For a masterly %%@
discussion, cf. Shimony (2004): his Sections 3 to 5 consider loopholes. Santos (2005, %%@
especially pp. 555-561) is a heterodox but admirably detailed discussion. I also set aside %%@
other loopholes, i.e. deniable implicit assumptions of Bell's theorem. Some are familiar, %%@
e.g. allowing some kind of backwards causation (Cramer 1986; Price 1996, Chapters 8,9; %%@
Berkovitz 2002, Section 5); others have only recently been articulated, such as the memory %%@
loophole and the collapse locality loophole (Kent 2005, and references therein). But I stress %%@
that these loopholes make it hard to secure a conclusive contradiction between quantum theory %%@
and ``local realism''; so that my blaming the Bell inequality's violation on outcome %%@
dependence, and in the sequel on the falsity of PCC and SEL, involves a judgment. Notoriously %%@
for philosophers, they also make it hard to infer spacelike causation from the Bell %%@
correlations (cf. Berkovitz 1998, 1998a, 2007 for detailed reviews; and Su\'{a}rez 2007); but %%@
I will not discuss causation, except {\em en passant} in connection with the PCC. I also set %%@
aside the conflict between quantum theory and {\em non}-local realism, i.e. the derivability %%@
of Bell inequalities from certain non-local stochastic hidden variable theories: cf. Fahmi %%@
and Golshani (2006), Socolovsky (2003) and Seevinck (forthcoming).}

The locality loophole is the idea that $\lambda$ is not (causally or stochastically) %%@
independent of (one or both of) the measurement choices, either because of causation between %%@
them or because both are influenced by a common cause. But most people regard the latter as %%@
an incredible conspiracy, and the former as ruled out by experiments in which the measurement %%@
choices are spacelike to the emission of the particles from the source.  I will concur with %%@
this (despite footnote 13).\footnote{The first such experiment was by Aspect et al. (1982); %%@
cf. also Weihs et al. (1998). Of course, `choice' here need not be a human decision: which %%@
quantity is measured might be determined by a random device; (for some details, cf. Shimony %%@
(2004, Section 5)).}

Note that in such an experiment each choice-event is also spacelike to a finite, albeit %%@
perhaps small, region comprising the ``summit'' of the past light-cone of the emission event. %%@
On account of these experiments it has become common, and it seems reasonable, for %%@
philosophical discussions of the fate of PCC in the Bell experiment to assume that the %%@
choice-events occur after any common cause event: we will return to this, especially in %%@
Section \ref{ccc}.

\subsubsection{My previous position}\label{prev}
 As I mentioned in Section \ref{Intr}, there are two motivations for exhibiting the  %%@
violation of SEL in the Bell experiment. First: our experience of the world, as codified in %%@
classical and relativistic physics, suggests that---in a slogan---causal processes cannot %%@
travel faster than light; and I submit that SEL is a plausible expression of that idea. So if %%@
we want to express precisely in which sense quantum theory is a ``non-local'' theory, or what %%@
is ``spooky'' about Bell correlations, the violation of SEL is a good candidate. In other %%@
words: it is worth exhibiting the violation of SEL, as a way of locating the mysteriousness %%@
of Bell correlations.

The second motivation is personal. In previous papers, I argued that outcome dependence %%@
implied that some formulations of SEL were violated in quantum mechanics. (This seemed worth %%@
showing since some previous work suggested the opposite.) I also endorsed the literature's %%@
folklore that outcome dependence implied the violation of the much more familiar condition, %%@
PCC (cf. Section \ref{comp}). Of course, these positions went hand in hand, in view of the %%@
similarities between PCC and the SELD formulations. To set the stage for later discussion, I %%@
need to recall these views: first about SEL, then about PCC. (Section \ref{causeaqft} will %%@
report my previous views about the fate of SEL in quantum field theory---and defend them!)

In my 1994, I gave three formulations of SEL. The first, called SEL1, was essentially our %%@
(Section \ref{3formul}'s) SELS. The second, called SEL2, was our SELD2. (The third used %%@
counterfactuals, and here I set it aside; our SELD1 was not discussed in (1994).) Then I %%@
argued that the Bell experiment in elementary quantum mechanics violated SEL (i.e. all %%@
formulations). (In this I disagreed with Hellman (1982), who hedged his corresponding %%@
formulations of SEL with provisos so as to secure that quantum mechanics, despite outcome %%@
dependence, obeyed them: details in Section \ref{3302}.)

In my (1989, 1992), I also joined other authors in holding that outcome dependence violated %%@
PCC. More precisely, I argued that what I took to be a weaker cousin of PCC  was violated. %%@
For PCC, in Section \ref{comp}'s form,  faces counterexamples that have nothing to do with %%@
the Bell inequality: everyday life and classical physics provide examples in which events $A, %%@
B$ and their common cause $C$---or the event that best deserves that name---do not satisfy %%@
eq. \ref{screen0}.

I argued that we should reply to these examples, by taking as the `common cause' of $E$ and %%@
$F$ the total physical state of a spacetime region that deserves (well enough, if not best) %%@
the name `common past of $E$ and $F$'.  In particular, it need not be the intersection of %%@
their past light-cones: in some cases, it needs to include parts of the boundaries of these %%@
light-cones, that are   future to the ``summit'' of $C^-(E) \cap C^-(F)$. (But it is in all %%@
cases a subset of the union $C^-(A) \cup C^-(B)$.) Taking $C$ as this total physical state, I %%@
dubbed eq. \ref{screen0} `PPSI' (for `Past Prescribes Stochastic Independence'). (Here, %%@
`total physical state' needs of course to be intrinsic; cf. Section \ref{prel}.)

 More precisely, I took this total physical state $C$ to prescribe a probability distribution %%@
$pr_C$, without itself being conditioned on; (Section \ref{Bellreview}'s many spaces %%@
approach). So PPSI said (1989, pp. 123-124):
\begin{quote}
{\bf PPSI}:
If spacelike events $E$ and $F$ are correlated but one does not cause the other, then:
\be
pr_C (E \& F) = pr_C (E).pr_C (F)
\label{ppsi}
\ee
where $pr_C$ is the  probability distribution prescribed by the total physical state $C$ of %%@
the common past of $E$ and $F$.
\end{quote}
I argued that by taking the common past as large as possible, and taking its {\em total} %%@
physical state, PPSI was weak, since one was in effect conditioning on more events---all %%@
those in the common past, both those that causally affect $E$ and $F$ and those that do not.

Then I argued that when we take $E$ and $F$ to be the measurement outcomes of a Bell %%@
experiment, the common past region could be chosen so as to satisfy three desiderata for %%@
proving Bell's theorem:\\
\indent(i): it is large, so that PPSI is weak (and plausible since satisfied by countless %%@
examples in everyday life and classical physics); \\
\indent(ii): PPSI justifies the outcome independence and parameter independence assumptions %%@
of Bell's theorem;\\
\indent(iii): the other assumptions of Bell's theorem, especially $\lambda$'s independence of %%@
measurement choices, are plausible; ({\em contra} the loopholes listed at the end of Section %%@
\ref{Bellreview}).\\
So my overall conclusion was that (as in my papers' discussion of SEL) the mysteriousness of %%@
the Bell correlations can be taken to lie in the violation of PPSI applied to measurement %%@
outcomes.\footnote{Cf. my (1989); especially pp. 122-130 for PPSI in general, and pp. 135-144 %%@
for the application to the Bell experiment, i.e. desiderata (i)-(iii). My (1992, pp. 74-76) %%@
is a summary.}

Two final comments about PPSI, which will be important  in Sections \ref{B} and %%@
\ref{SELBE}.\\
\indent (1): In taking PPSI to be plausible---especially, a reply to the everyday %%@
counterexamples to PCC---I assume that `conditionalizing on all the other [i.e. causally %%@
irrelevant] events in the [common past] does not disrupt the stochastic independence induced %%@
by conditionalizing on the affecting events' (1989, p. 123). That conditionalizing can %%@
produce stochastic dependence is sometimes called `Simpson's paradox'; so my position assumes  %%@
that Simpson's paradox does not apply. We shall return to this assumption, especially in %%@
Section \ref{ccc}.\\
\indent (2): PPSI is like Section \ref{3formul}'s formulations of SEL (and Bell's `local %%@
causality') in having the total state of a spacetime region prescribe a probability %%@
distribution, rather than be an event that is conditionalized on (as in Reichenbach's PCC, %%@
eq. \ref{screen0}). In Section \ref{Bellreview}'s terms: PPSI adopts the many spaces %%@
approach. This has two important consequences, which will be developed below.\\
\indent \indent (2a): The similarity between PPSI and SEL makes it easier work to argue from %%@
the Bell experiment's violation of PPSI, that it also violates SEL (Section \ref{SELBE}).\\
\indent \indent (2b): Conditionalizing on an event as in eq. \ref{screen0} is easily %%@
generalized to considering a partition of the probability space, and conditionalizing on each %%@
of its cells. And if one is considering several pairs of correlated events,  one can ask %%@
whether (i) there is a single partition (each cell of) which yields a screening-off, or (ii) %%@
there are several partitions, one for each pair of correlated events. This will be the theme %%@
of Section \ref{B}. But this contrast can hardly be expressed if, as in  PPSI and SEL, we %%@
adopt the many spaces approach and ``go all the way down'' to an individual world, or history %%@
up to a hypersurface, or total physical state of a common past, which we then take to %%@
prescribe a probability distribution. We will return to this feature of PPSI and SEL in %%@
Sections \ref{formloc} and \ref{SELBE}.

\subsection{A {\em common} common cause? The Budapest school}\label{B}
So much by way of recalling my previous views. But these now need to be re-examined in the %%@
light of the subsequent literature. The first thing to say is that happily, some later %%@
authors endorse them. For example, Uffink and Henson join me in following the lead of Bell's %%@
local causality, but allowing the common past to include more than the intersection of the %%@
past light-cones. So they formulate a condition very like my PPSI, and then say that outcome %%@
dependence violates it; (Uffink 1999, pp. S523-S524; Henson 2005, pp. 516-533).

But as I said in Section \ref{Intr}, the main development I need to address is the Budapest %%@
school's work objecting to the folklore that outcome dependence violates PCC, on the grounds %%@
that PCC does not require different correlations to have the same common cause (a `common %%@
common cause'). I will discuss this work (Section \ref{resus} to \ref{sz}), but then %%@
maintain, following ``replies'' by Placek and the Bern school, that the folklore is right: %%@
outcome dependence violates PCC (Section \ref{ccc} and \ref{avoid}). (So much of Sections %%@
\ref{resus} to \ref{avoid} is expository.)  Then in Section \ref{SELBE}, I will carry the %%@
discussion  over to PPSI and SEL.\footnote{For my purposes in this Section, the main %%@
references for the Budapest school are: Hofer-Szab\'{o} et al. (1999, 2002), Szab\'{o} %%@
(2000), R\'{e}dei (2002) and Hofer-Szab\'{o} (2007); and the main references for the Bern %%@
school are: Grasshoff et al. (2005) and Portmann and W\"{u}thrich (2007). Placek is of course %%@
a leader of the (equally prolific!) ``Pittsburgh-Krakow school'': they develop a rigorous %%@
framework combining modality (indeterminism, ``branching spacetime''), events' spacetime %%@
locations, and their probabilities---and then use this framework to analyse the Bell %%@
experiment. This framework is fruitful: for example, its branching structure secures an %%@
algebraic (Galois connection) definition of the different outcomes of a stochastic event, and %%@
a theorem that they form a Boolean algebra (Kowalski and Placek 1999, Section 2). It also %%@
provides rigorous formulations of: (i) two incompatible measurements or outcomes occurring in %%@
the same spacetime region; (ii) the PCC in both weak and strong versions; and (iii) Bell %%@
theorems. Important recent references include: Placek 2002, 2004, Belnap 2005, M\"{u}ller %%@
2005.  Besides, SEL could surely be formulated in it (even more rigorously than in Section %%@
\ref{formul}!), so as to assess whether SEL  is violated by outcome dependence. But I will %%@
duck out of this project, since the framework is so rigorous as to make it a considerable %%@
effort: and I wager that it would make no difference to my conclusions. In particular, we %%@
will see that Placek's defence of a common common cause can be separated from the framework.}

\subsubsection{Resuscitating the PCC}\label{resus}
The Budapest school provides two kinds of resuscitation of PCC: formal, and physical. The %%@
formal resuscitation consists of rigorous theorems that under certain conditions, common %%@
cause events---technically: events in a probability space that screen off---must exist. There %%@
are such theorems, both for classical and for quantum probability spaces; and also in the %%@
framework of AQFT. But here I only need a rough statement of one main theorem for classical %%@
probability. The idea is to proceed in two steps.\\
\indent (1): Given a probability space, say $S$, and a pair of correlated events $E, F$ (so %%@
$E, F \subset S$), $S$ may well not contain a common cause in Reichenbach's sense. But one %%@
can build another probability space $\bar{S}$, with the features: (i) $S$ can be mapped  %%@
one-to-one into $\bar{S}$, while preserving both the algebraic (set-theoretic) relations %%@
between, and the probabilities of, events; call this map $h$; (ii) $\bar{S}$ contains a %%@
common cause of $h(E)$ and $h(F)$. In fact, $\bar{S}$ is built from the disjoint union of two %%@
copies of the given space $S$.\\
\indent  (2): Given any finite set of pairs of correlated events in a probability space, one %%@
can pick one pair after another, iterating the construction in step (1); and so conclude by %%@
induction that for any such finite set of pairs of correlated events, there is a probability %%@
space which for each pair contains a Reichenbachian common cause.\footnote{For details of (1) %%@
and (2), cf. Proposition 2 of Hofer-Szab\'{o} et al. (1999). The Pittsburgh-Krakow school has %%@
a theorem with a similar flavour; cf. Placek (2000,pp. 456-459 ; 2000a, pp. 176-178; 2002, %%@
pp. 333-334). The Budapest school also proves a restrictive necessary condition on the %%@
probabilities of two pairs of correlated events, for them to have a common common %%@
cause---even in an extension of $\bar{S}$ of the given  space $S$; (Hofer-Szab\'{o} et al. %%@
(2002), Proposition 4). This restriction prompts the next paragraph's discussion of the %%@
strong PCC.}

The physical resuscitation is based on distinguishing, for situations with more than one pair %%@
of correlated events,  a weak PCC and a strong one. The weak PCC requires only that for each %%@
pair of correlated events, there is a screener-off, with different pairs in general having %%@
different screeners-off. This weak version  seems to have been Reichenbach's main idea. In %%@
any case, it is certainly this weak version that is vindicated by the theorem just reported: %%@
for $n$ such pairs, each of step (2)'s $n$ iterations of (1)'s construction  will define a %%@
new event (a subset of the new disjoint union) as the common cause of the pair being %%@
addressed at that iteration.
On the other hand, the strong PCC requires that all pairs have the same screener-off: a {\em %%@
common} common cause. So the strong PCC is not vindicated by the above theorem.\footnote{Nor %%@
could it be, in view of the necessary condition mentioned in the previous footnote. Nor, %%@
incidentally, is the strong PCC vindicated by the theorem's quantum analogue; to check this, %%@
cf. eq (36) et seq. of Proposition 3 of Hofer-Szab\'{o} et al. (1999). But I shall not need %%@
to consider the Budapest school's quantum version of the PCC, in either weak or strong %%@
versions: which it would be worthwhile to compare with Henson's version (2005, pp. 534-536). %%@
Nor, incidentally, need I consider Uffink's (convincing!) proposal for how to extend the PCC %%@
to multiple events (1999, pp. S517-S520).}

Let us first make this distinction more precise, while also generalizing to allow the common %%@
cause to be, not a dichotomic event that either occurs or not ($C$ or $\neg C$), but---as %%@
needed for the Bell experiment---the value of a variable, with maybe more than two values. On %%@
analogy with eq. \ref{facby} et seq., and Reichenbach's own requirement that both the common %%@
cause $C$ and its negation $\neg C$ screen off $E$ and $F$, we shall require stochastic %%@
independence at each value of the variable, $\lambda$ say, in its range $\Lambda$. So in %%@
discussing common causes for a general probability space $S$, $\Lambda$ corresponds to a %%@
partition of $S$, and a value $\lambda \in \Lambda$ to a cell. Of course, an example of such %%@
a space $S$ appropriate to the Bell experiment is the ``big space'' whose elements are %%@
quintuples $<\lambda, x, y, X, Y>$, mentioned in Section \ref{Bellreview}.\footnote{On the %%@
other hand, we can keep things simple by taking the correlated events themselves always to be %%@
dichotomic; (essentially because the quantities at issue in the Bell experiment are %%@
two-valued). A note on jargon: the Budapest school calls cases where the common cause is a %%@
cell of a partition (value of a variable) rather than a single event `common cause systems'; %%@
and so a strong PCC requires a `common common cause system'. Cf. Hofer-Szab\'{o} (2007, %%@
Sections 1 and 2) for a review of the definitions, and references.}

Then a weak PCC, and a strong PCC, can be stated as follows. Let $\{E_m \}, \{F_m \}$ be two %%@
sets of events in a probability space $S$. We say the set of (ordered) pairs $(E_m, F_m)$ is %%@
{\em correlated} if for some (maybe all) values of $m$, $pr(E_m \& F_m) \neq pr(E_m) \cdot %%@
pr(F_m)$. Then the weak PCC states: for each value of $m$ for which there is a correlation, %%@
there is a partition, $\Lambda^m$ say, of the probability space $S$ (i.e. there is a common %%@
cause variable) such that for every value (cell) $\lambda^m_l$ of $\Lambda^m$:
\be
pr(E_m \& F_m / \lambda^m_l) = pr(E_m / \lambda^m_l) \cdot pr(F_m / \lambda^m_l).
\label{screenweak}
\ee
(We assume that the relevant conditional probabilities are non-zero; cf. footnote 6. The %%@
range of $l$, i.e. the number of cells in the partition, may of course depend on $m$.) Again: %%@
the stochastic independence at {\em each} value (cell) $\lambda^m_l$ generalizes %%@
Reichenbach's requiring screening-off both by $C$ and by $\neg C$.

On the other hand, the strong PCC reverses the quantifiers to give a stronger `there exists, %%@
for all' statement: viz., there is a partition (a common cause variable), $\Lambda$, of the %%@
probability space $S$ such that, for each value of $m$ for which the pair $(E_m, F_m)$ is %%@
correlated, and for every value $\lambda$ of $\Lambda$:
\be
pr(E_m \& F_m / \lambda) = pr(E_m / \lambda) \cdot pr(F_m / \lambda).
\label{screenstrong}
\ee
Clearly, this single partition $\Lambda$ of $S$ deserves the name `common common cause' (or %%@
`common common cause variable').

Since the Bell experiment involves several pairs of correlated events, this distinction %%@
applies to it.  (For most quantum states and measurement choices: all four joint measurement %%@
choices give a correlation, so that since all four quantities are two-valued, there are four %%@
pairs of correlated events.) The Budapest school now claims that proofs of a Bell inequality %%@
seem to always use a strong version of PCC; i.e. to invoke a common common cause, along the %%@
lines of eq. \ref{screenstrong}.\footnote{This point was first made by Belnap and Szab\'{o} %%@
(1996); cf. also e.g. Hofer-Szab\'{o}, G.  et al. (1999), p. 388.}  If so, then the violation %%@
of the Bell inequality impugns at worst the strong PCC---not the weak one, nor therefore  %%@
Reichenbach's original formulation.

 This last claim prompts five discussions, taken up in the following Subsubsections. The %%@
first and third  support the claim; the second clarifies it by stating two distinctions about %%@
how to think of the space $\Lambda$. Then the fourth and fifth reply to the claim, in two %%@
ways. These replies follow Placek and the Bern school, respectively;  (adding to their %%@
discussions only the deployment of the two distinctions). Then with these discussions about %%@
PCC in hand, Section \ref{SELBE} will be able to make quick work of SEL.

\subsubsection{Known proofs of a Bell inequality need a strong PCC}\label{need}
First, the claim is plausible, when we consider how known proofs of a Bell inequality %%@
proceed: compare eq. \ref{screenstrong} with eq. \ref{facby} to \ref{oi}. More generally, we %%@
notice that:\\
\indent (a):  Bell inequalities are linear inequalities involving joint probabilities (and in %%@
some versions, also: single probabilities) of outcomes. \\
\indent (b): A hidden variable model of the experiment represents each of these probabilities %%@
as a weighted average (convex combination) of ``hidden'' conditional probabilities for the %%@
joint (or single) outcome, viz. the probabilities conditional on each value of  $\lambda$. \\
\indent (c): In known derivations of the Bell inequality, it is crucial that the same %%@
variable $\lambda$, and so the same weights $pr(\lambda)$, be used in each term of the %%@
inequality; and that the model is ``local'' in that this $\lambda$ provides a strong PCC %%@
along the lines of eq. \ref{screenstrong}. For it is only with these assumptions that we can %%@
argue: first, that the  joint (and maybe single) probabilities conditional on each value of  %%@
$\lambda$ obey an inequality (as a matter of arithmetic); and then that averaging over the %%@
values of $\lambda$---and so going from hidden conditional probabilities to observable %%@
probabilities---preserves the inequality.

 One can also support the claim, other than by surveying proofs. Namely, one can try to build %%@
a  hidden variable model of the Bell experiment that is empirically adequate (i.e. delivers %%@
the correct quantum correlations) and uses only the {\em weak} PCC, and is otherwise %%@
physically reasonable. Here one naturally takes `physically reasonable' to mean analogues of %%@
Section \ref{Bellreview}'s conditions. Thus one seeks a model in which for each postulated %%@
common cause variable, $\lambda^m$:\\
\indent (i) appropriate versions of both parameter independence and outcome independence hold %%@
good; and\\
\indent (ii) $\lambda^m$ is not correlated with the choice of measurement.

In Section \ref{sz}, I will describe how Szab\'{o} (2000) builds such a model; (he adapts the %%@
disjoint union construction reported in (1) at the start of Section \ref{resus}). But the %%@
exact formulation of conditions (i) and (ii) needs some care. For the formulation  depends %%@
on:\\
\indent (Many-Big): the distinction  mentioned in Section \ref{Bellreview}, between the many %%@
spaces and the  big space approach to formulating Bell's theorem;\\
\indent (Only-Total): the distinction  mentioned in Section \ref{prev}, about whether or not %%@
$\lambda$ encodes only factors that are causally relevant to the outcomes.\\
\indent These distinctions will also play a role in our eventual reply to the Budapest %%@
school's claim. So I turn to filling out these distinctions, (Many-Big) and (Only-Total).

\subsubsection{Two distinctions}\label{two}
\paragraph{The distinctions stated}\label{stated}
 As to (Many-Big), we saw in Section \ref{Bellreview} that:\\
\indent\indent  (i): The many spaces approach postulates in the first instance a space %%@
$\Lambda$ of hidden variables; one may then build other spaces from that, e.g. by taking a %%@
Cartesian product. On the other hand;\\
\indent\indent  (ii): The big space approach takes a value $\lambda$ to be an event (i.e. %%@
subset not an individual element) in a larger space that also has  measurement choices and %%@
outcomes as events.

Then in Section \ref{prev} we stated (implicitly) another distinction, (Only-Total), which %%@
cuts across (Many-Big). Namely, the distinction whether \\
\indent \indent  (i): $\lambda$ encodes some kind of intrinsic state of the particle-pair at %%@
the time of emission, or maybe a bit later; (but certainly before they enter the apparatuses, %%@
so that the distribution $\rho$ can be independent of the measurement choices $x$ and $y$); %%@
or \\
\indent \indent (ii): $\lambda$ encodes the total physical state of a spacetime region (maybe %%@
with sub-regions lying to the future of the emission event); $\lambda$ is thereby myriadly %%@
complex and no doubt contains many features that are causally, indeed stochastically, %%@
irrelevant to the outcomes. (As we saw, this was the option taken by SEL and PPSI, though %%@
without the  $\lambda$ notation.)

This distinction (Only-Total) cuts across the first one (Many-Big), since, for example, the %%@
total physical state in (ii) of (Only-Total) need not encode anything about measurement %%@
choices and outcomes. Indeed, it should not do so, so as to better support Section %%@
\ref{Bellreview}'s ``locality'' assumptions of a Bell's theorem.

 In Section \ref{resus} we thought of a common cause variable as a partition $\Lambda^m$ of %%@
the probability space $S$, with a common cause event $\lambda^m_l$ being a cell of the %%@
partition (a value of the variable). Since in general a cell is not a  singleton set, there %%@
will be more than one way a given value $\lambda^m_l$ can be realized. Clearly, this can be %%@
motivated by adopting the second option, (ii), for each (or both) of the distinctions %%@
(Many-Big) and (Only-Total).\\
\indent As to (ii) of (Many-Big): if the {\em elements} of the probability space $S$ encode %%@
measurement choices and outcomes, then there will certainly be more than one way that a  %%@
given value $\lambda^m_l$ can be realized. For the simple space of quintuples %%@
$<\lambda,x,y,X,Y>$ with which I introduced the big space approach, there will be 4 $\times$ %%@
4 = 16 ways, since all four joint measurement choices can have four results.\\
\indent As to (ii) of (Only-Total): if $\lambda$ encodes the total physical state of a %%@
spacetime region, it will be very logically strong (``rich'') and will in general include %%@
countless features that are causally, indeed stochastically, irrelevant to the outcome. So %%@
there will be countless ways that a given value of a common cause variable $\lambda^m_l$ can %%@
be realized. Note here that we firmly distinguish $\lambda^m_l$, a cell of a partition of %%@
$S$, from this rich $\lambda$, which is: either  an element of $S \equiv \Lambda$ (viz. on %%@
the many spaces approach), or a component of such an element (viz. on the big space %%@
approach).

With these two distinctions in hand, I can now make good the promise at the end of Section %%@
\ref{need}, to formulate some appropriate versions of  parameter independence, outcome %%@
independence, and the condition that the common cause is not correlated with the choice of %%@
measurement.

\paragraph{Formulating locality conditions}\label{formloc}
Though it would be a good exercise to write such formulations for all four cases arising from %%@
the two distinctions, I shall here consider just one case, viz. the many spaces approach, and %%@
take $\lambda$ to encode a total physical state of a region; (i.e. options (i) of (Many-Big) %%@
and (ii) of (Only-Total)). We shall also see in Section \ref{sz} that Szab\'{o} treats %%@
another of the four cases: he adopts the big space approach, and (implicitly) option (i) of %%@
(Only-Total). These two cases---ours here, and Szab\'{o}'s in Section \ref{sz}---will be %%@
enough for our goals, namely  replying to the Budapest school's claim about Bell theorems %%@
(Sections \ref{avoid} and \ref{ccc}); and arguing that SEL fails in the Bell experiment %%@
(Section \ref{SELBE}).

So let us for now adopt the many spaces approach, and take $\lambda$ to encode a total %%@
physical state of a region. Then $S \equiv \Lambda \ni \lambda$ and $\lambda^m_l$ is a cell %%@
of a partition $\Lambda^m$ of $S$. Let us consider in order, formulating:\\
\indent (i) parameter independence,\\
\indent (ii) outcome independence,\\
\indent (iii) putting these together: factorizability; and \\
\indent (iv) the common cause being  independent of the measurement choice.\\
We will see that for each of these, we have a choice: we can either:\\
\indent (a) retain Section \ref{Bellreview}'s original formulations, or \\
\indent (b) incorporate a reference to a (in general: non-common) common cause $\lambda^m$.\\
On either option, we get an equivalence between a conjunction of (i) and (ii), with (iii). %%@
And both options will be relevant to our reply to the Budapest school: the first option will %%@
be taken up in Section \ref{ccc}, the second in Section \ref{avoid}.

(i): {\em Parameter independence}:--- Option (a): We can repeat  eq. \ref{pi}, word for word. %%@
That is, we can say that each ``rich micro-state'' $\lambda$ screens off a nearby outcome and %%@
a distant setting, without mentioning common cause variables $\lambda^m$. Thus, we repeat eq. %%@
\ref{pi} for L outcomes:
\be
\forall \lambda; x,y; X, Y = \pm 1: \;\;
pr_{\lambda, x}(X) = pr_{\lambda, x,y}(X) :=   pr_{\lambda, x,y}(X \& Y) + pr_{\lambda, %%@
x,y}(X \& \neg Y) \; .
\label{pi2}
\ee
Or Option (b): we postulate that there are probability functions $pr_x, pr_y, pr_{x,y}$, and
require that for each joint setting $m = <x,y> = <a_1,b_1>,...,<a_2,b_2>$, there is a %%@
partition $\Lambda^m$ of $\Lambda$, of which each cell $\lambda^m_l$ screens off each nearby %%@
outcome and its distant setting. Besides, since $\lambda^m_l \subset \Lambda$ is an event, we %%@
express this screening-off with a conditional probability, not a subscript---despite having %%@
adopted the many spaces approach. That is, we require that $pr_x, pr_y, pr_{x,y}$ satisfy:
\begin{eqnarray}
\label{picoarse}
\forall \; m \; = \; <x,y> \; = \; <a_1,b_1>,...,<a_2,b_2>; \;\; \exists \Lambda^m , \;\; %%@
\forall \; {\rm{cells}} \; \lambda^m_l; \;\; \forall X, Y = \pm 1: \\ \nonumber
pr_{x}(X / \lambda^m_l) = pr_{x,y}(X / \lambda^m_l) :=   pr_{x,y}(X \& Y  / \lambda^m_l) + %%@
pr_{x,y}(X \& \neg Y / \lambda^m_l) \; ;
\end{eqnarray}
(and similarly for R outcome probabilities).

Neither formulation implies the other. Obviously, eq. \ref{pi2} does not imply eq. %%@
\ref{picoarse}, since ``coarse-graining'' (``losing information'') can destroy screening-off. %%@
(To be precise: if we read eq. \ref{picoarse} as having a fixed partition, then eq. \ref{pi2} %%@
does not imply eq. \ref{picoarse}. Of course, eq. \ref{pi2} {\em does} imply eq. %%@
\ref{picoarse} read as asserting that there is some such partition: for eq. \ref{pi2} {\em %%@
is} eq. \ref{picoarse} with the finest possible partition, given by the singletons of all the %%@
elements $\lambda \in \Lambda$.)

 Conversely, eq. \ref{picoarse} does not imply eq. \ref{pi2}, since``fine-graining'' i.e. %%@
conditionalizing can produce stochastic  dependence (`Simpson's paradox'; cf. (1) at the end %%@
of Section \ref{prev}).

(ii): {\em Outcome independence}:--- The situation  is entirely parallel. Option (a): We can %%@
repeat  eq. \ref{oi}, word for word, saying that each  $\lambda$ and joint setting screens %%@
off the two outcomes, without mentioning common cause variables $\lambda^m$. Thus, we repeat %%@
eq. \ref{oi}:
\be
\forall \lambda, x,y; X, Y = \pm 1: \;\;
pr_{\lambda, x,y}(X \& Y) =  pr_{\lambda, x,y}(X) \cdot pr_{\lambda, x,y}(Y) \; .
\label{oi2}
\ee
Or Option (b): We can require that for each joint setting $m = <x,y>$, there is a partition %%@
$\Lambda^m$ of $\Lambda$, of which each cell $\lambda^m_l$ screens off the two outcomes. And %%@
as in eq. \ref{picoarse}, we express the screening-off by conditionalizing on the event %%@
$\lambda^m_l$---despite having adopted the many spaces approach. That is, we require:
\begin{eqnarray}
\label{oicoarse}
\forall \; m \; = \; <x,y> \; = \; <a_1,b_1>,...,<a_2,b_2>; \;\; \exists \Lambda^m , \;\; %%@
\forall \; {\rm{cells}} \; \lambda^m_l; \;\; \forall X, Y = \pm 1: \\ \nonumber
pr_{x,y}(X \& Y / \lambda^m_l) = pr_{x,y}(X / \lambda^m_l) \cdot pr_{x,y}(Y / \lambda^m_l) \; %%@
.
\end{eqnarray}

As with parameter independence in (i), neither formulation implies the other---and for %%@
parallel reasons: both ``coarse-graining''  and ``fine-graining'' can destroy stochastic %%@
independence.  We will return to the converse non-implication---that ``fine-graining'' can %%@
destroy screening-off, i.e. Simpson's paradox---in Section \ref{ccc}.

(iii): {\em Factorizability}:--- Recall from Section \ref{Bellreview} that eq. \ref{facby} is %%@
equivalent to the conjunction of eq.s \ref{pi} and \ref{oi}. The equivalence generalizes %%@
easily to the coarse-grained versions considered here, provided each measurement choice %%@
determines the same partition for all three conditions. That is to say: one easily checks %%@
that coarse-grained factorizability, viz.
\begin{eqnarray}
\label{facbycoarse}
\forall \; m \; = \; <x,y> \; = \; <a_1,b_1>,...,<a_2,b_2>; \;\; \exists \Lambda^m , \;\; %%@
\forall \; {\rm{cells}} \; \lambda^m_l; \;\; \forall X, Y = \pm 1: \\ \nonumber
pr_{x,y}(X \& Y / \lambda^m_l) = pr_{x}(X / \lambda^m_l) \cdot pr_{y}(Y / \lambda^m_l) \; .
\end{eqnarray}
is equivalent, {\em not} to the conjunction of eq.s \ref{picoarse} and \ref{oicoarse} as they %%@
stand, each saying `there is a partition $\Lambda^m$': but to the stronger statement with the %%@
conjunction in the scope of the existential quantification: i.e. the statement that for any %%@
$m$, there is a (single) partition $\Lambda^m$ all of whose cells do both sorts of screening %%@
off---outcomes from distant settings, as in eq. \ref{picoarse}, and outcomes from each other %%@
as in eq. \ref{oicoarse}.

(iv): {\em Independence of measurement choice}:--- We have a corresponding choice about how %%@
to formulate the idea that the common cause is independent of the measurement choice. First, %%@
Option (a): we can take `common cause' as in Section \ref{Bellreview} to mean the ``rich %%@
micro-state'' $\lambda \in \Lambda$: then the independence is expressed exactly as before, by %%@
the postulation of a fixed probability measure $\rho$ on $\Lambda$ (rather than %%@
$\rho_{x,y}$).

Or Option (b): we can take `common cause' to mean a cell (event) $\lambda^m_l$ of a partition %%@
$\Lambda^m$ defined by a measurement choice. In that case, there is a sense in which the %%@
common cause is trivially {\em dependent} of the measurement-choice. Namely: different values %%@
of $m \; = \; <x,y>$ will in general  define different events $\lambda^m_l$ (even if we pick %%@
the same cell label $l = <\pm 1, \pm 1>$ out of the four possibilities), with in general %%@
different probabilities---despite there being a fixed measure $\rho$.\footnote{But as I %%@
mentioned at the end of Section \ref{need}, Szab\'{o}'s model will obey a substantive %%@
condition requiring a non-common common cause to be stochastically independent of the %%@
measurement choice. For Szab\'{o} adopts the big space approach, and so requires an analogue %%@
of Section \ref{Bellreview}'s eq. \ref{lambchoice}.}

Note that if in each of (i) to (iv) we take Option (a), i.e. we retain the formulation given %%@
in Section \ref{Bellreview}, then we will be committed to a Bell inequality:  despite our %%@
education in Budapest, so to speak---that in general, common causes are not common common %%@
causes. We will return to this in Section \ref{ccc}'s reply to Budapest, and in Section %%@
\ref{SELBE}.

On the other hand, if in each of (i) to (iv) we take  Option (b), i.e. we assume only %%@
$m$-dependent common cause variables, then the Budapest school's lessons (from Section %%@
\ref{need}, and to come, from Section \ref{sz}) apply. That is: from these lessons, it seems %%@
that a Bell inequality cannot be derived. But beware: we will return to this option in %%@
Section \ref{avoid}.

So much about how to formulate locality conditions when we adopt option (i) of (Many-Big) and %%@
option (ii) of (Only-Total). I turn to Szab\'{o}'s many-common-cause  model of Bell %%@
correlations, promised at the end of Section \ref{need}.

\subsubsection{Szab\'{o}'s model}\label{sz}
Szab\'{o}'s model adopts the second option of (Many-Big). That is, he adopts a big space %%@
approach. But his space is considerably more complex than the space of quintuples %%@
$<\lambda,x,y,X,Y>$ with which I introduced this approach. For he constructs his model's %%@
common cause events by adapting Hofer-Szab\'{o}, G.  et al. (1999)'s iterated construction %%@
that takes a disjoint union of two copies of the probability space; (cf. (1) and (2) in %%@
Section \ref{resus}).

For the sake of completeness, I should add about the distinction (Only-Total), that Szab\'{o} %%@
is of course aware of it: his Section 14 sketches option (ii) of (Only-Total), in which %%@
$\lambda$ encodes a total physical state. But he does not adopt this option. In fact, for %%@
each of the four correlations, the common cause that Szab\'{o} constructs at each stage is %%@
one of the two copies of the given probability space, i.e. one of the ``unionands'' of the %%@
disjoint union: details below.

I can now state the locality conditions obeyed by Szab\'{o}'s model, with its  big space %%@
approach.
 I shall present them in the order in which Section \ref{Bellreview} introduced them, adding %%@
a reference to Szab\'{o}'s equations; (our notations are similar).

The first point to make  concerns the  probabilities of the elements of his probability space %%@
(the atoms of his event algebra). As  Szab\'{o} himself stresses (2000, Sections 13-14), %%@
these probabilities depend on the probabilities of measurement choices; {\em contra} our %%@
subscript-less $\rho$ in Section \ref{Bellreview}. But (as I announced in footnote 11) this %%@
is {\em not} suspicious: it in no sense suggests a ``conspiracy''. For on Szab\'{o}'s big %%@
space approach, each element encodes the entire experiment; or in logical language: it %%@
represents a truth-value assignment to propositions reporting choices and outcomes, as well %%@
as those reporting the values of common cause variables, i.e. which common cause events %%@
occur. In short, this dependence is innocuous: just as it is for the probabilities of the %%@
quintuples $<\lambda,x,y,X,Y>$ in my ``baby'' example of the big space approach.

This means that our requirement that the measurement choices be independent of the hidden %%@
variable $\lambda$ (our eq. \ref{lambchoice}) goes over in Szab\'{o}'s model to a requirement %%@
that choices be stochastically independent of the common cause events; where each %%@
correlation, labelled $(X,Y)$ ($X= A_1, A_2; Y = B_1, B_2$) has its own common cause event, %%@
written $Z_{XY}$. So  Szab\'{o}'s eq. (13) is:
\be
p(x \&  Z_{XY}) = p(x)p(Z_{XY}) \;\;\; ; \;\;\; p(y \&  Z_{XY}) = p(y)p(Z_{XY}) \; .
\label{szlambchoice}
\ee

Parameter independence (our eq. \ref{pi}) is expressed by the stochastic independence of a %%@
nearby outcome and a distant setting. Thus Szab\'{o}'s eq. (12) is:
\be
p(X \& y) = p(X)p(y) \;\;\; ; \;\;\; p(Y \& x) = p(Y)p(x) \; .
\label{szpi}
\ee

Outcome independence (our eq. \ref{oi}) is now weakened to allow each correlation $(X,Y)$  to %%@
have its own common cause event $Z_{XY}$. But corresponding to eq. \ref{oi}'s requirement %%@
that all values of $\lambda$ screen-off the outcomes, Szab\'{o} requires that $\neg Z_{XY}$, %%@
as well as $Z_{XY}$, screens off. So  Szab\'{o}'s eq.s (17), (18) are:
\be
p(X \& Y / Z_{XY}) = p(X / Z_{XY})p(Y / Z_{XY}) \; ; \;
p(X \& Y / \neg Z_{XY}) = p(X / \neg Z_{XY})p(Y / \neg Z_{XY}) .
\label{szoi}
\ee

Finally, Szab\'{o} requires that the measurement choices in each wing are independent of each %%@
other. Thus his eq. (11) is
\be
p(x \& y) = p(x)p(y) \; .\footnote{Though I said `Finally', Szab\'{o} also imposes some other %%@
requirements on common causes, inspired by the Reichenbach tradition; cf. his eq.s (14), %%@
(16), (19). For example, they concern whether common causes promote or inhibit; (cf. (ii) in %%@
footnote 8). But we can ignore these extra requirements.}
\label{szindchoice}
\ee

So much for Szab\'{o}'s requirements on his models. I turn to a sketch of how he constructs %%@
his common causes, so as to make vivid how each correlation gets a different common cause. As %%@
I mentioned, he iterates the disjoint union construction four times, in each case %%@
constructing a common cause. In fact, the common cause is in each case just one of the two %%@
copies of the given space. Thus he starts with a probability space $\Omega$ with 16 elements %%@
$u_1,...,u_{16}$: and he identifies various sets of the $u$s with the four measurement %%@
choices and the four outcomes.  Then he considers two copies of this space: $\Omega_i := \{ %%@
(u,i) : u \in \Omega \}, i = 1,2$, and the 32-element probability space $\Omega' := \Omega_1 %%@
\cup \Omega_2$. He shows that he can define the common cause event of one of the %%@
correlations, say $(A_1,B_1)$, as one of the copies, say $\Omega_1$. Then at the next stage %%@
he takes two copies $\Omega'_i$ of $\Omega'$, and defines the 64-element probability space as %%@
the disjoint union $\Omega'' := \Omega'_1 \cup \Omega'_2$. Now he defines the common cause %%@
event of a second of the correlations, say $(A_1,B_2)$, as one of the copies, say %%@
$\Omega'_1$, considered as a subset of $\Omega''$. And so on, up to defining his final %%@
256-element probability space $\Omega'''' := \Omega'''_1 \cup \Omega'''_2$, and taking the %%@
common cause of the fourth correlation, say $(A_2,B_2)$, as $\Omega'''_1$. This makes it %%@
vivid how the four common cause events are distinct:  they have, respectively, 16, 32, 64 and %%@
128 atoms!

To sum up: Szab\'{o}'s model is impressive, both in its iterated technical construction, and %%@
in the range of requirements (eq. \ref{szlambchoice} to \ref{szindchoice}) it satisfies. He %%@
is surely right to conclude that the `model satisfies all locality conditions [that have] %%@
been required in the EPR-Bell literature'.

 But there is a catch, duly acknowledged by Szab\'{o}. Though each common cause event is %%@
independent of the measurement choices (eq. \ref{szlambchoice}), various Boolean combinations %%@
of these events, such as $Z_{A_1B_1} \cap Z_{A_1B_2}$ and $Z_{A_1B_1} \cup Z_{A_1B_2}$ {\em %%@
are} correlated with the measurement choices. This leads to the Bern school's ``reply'' to %%@
the Budapest school: cf. Section \ref{avoid}. But first we develop another, less technical, %%@
reply. 

\subsubsection{A common common cause is plausible}\label{ccc}
A good case can be made that in the Bell experiment, it is physically reasonable to postulate %%@
a common common cause. That is: a good case can be made for Section \ref{sofar}'s folklore %%@
that what is ``spooky'' about Bell correlations is that outcome dependence violates the %%@
strong PCC. I shall state this case; and then admit that, as the jokes say, there is bad news %%@
as well as good news.\footnote{I think that before the Budapest school's critique, I and most %%@
other ``folk'' had this case for common common causes ``unconsciously in mind''; and more %%@
recently, Uffink (1999, pp. S517, S523) and Henson (2005, pp. 527, 530) surely have it in %%@
mind. But all credit to Placek, who was, so far as I know, the first person to articulate %%@
this sort of case, in the aftermath of the Budapest school's critique (2000, pp. 464-465; %%@
2000a p. 185, 187). I also think that both the good and the bad news should be %%@
uncontroversial; for example, Grasshoff et al. (2005, p. 678) endorse both.}

The main idea why a common common cause is reasonable is straightforward.  Any two runs of a %%@
(well-designed!) Bell experiment have in common a great many features that are obviously %%@
relevant to the outcomes obtained (and resulting correlations) {\em apart from} the %%@
measurement choices: the preparation of the source and the two channels to the wings, the %%@
preparation of the detectors apart from their settings, the quantum state of the emitted %%@
particle-pair. (Agreed, any two runs will also differ in countless microscopic details about %%@
(a) these features and (b) other respects.) So it is reasonable to take these common features %%@
to define a hidden variable $\lambda$ which is common for the different measurement choices.

This case can be made stronger (or at least: more precise!) by invoking the assumption that %%@
we can arrange that the measurement choices are:\\
\indent (i): stochastically independent of these common features (i.e. of any common cause %%@
event); \\
\indent (ii): made after these features are settled.\\
Again, this assumption is reasonable. For (i) is accepted not just by the usual models (e.g. %%@
eq. \ref{lambchoice}), but also by Szab\'{o}'s model (eq. \ref{szlambchoice}). And (ii) is %%@
reasonable in the light of the Aspect experiment, i.e. the fact that we can arrange each %%@
measurement choice to be spacelike to a finite, if small, region comprising the ``summit'' of %%@
the past light-cone of the emission event.\\
\indent If we accept  this assumption, then having the hidden variable (the common features) %%@
be different for different measurement choices seems to be backward causation.

This case is also illustrated by the discussion in Section \ref{formloc}. There, we %%@
considered two cross-cutting distinctions, (Many-Big) and (Only-Total). We saw that if we %%@
adopt the many spaces approach, and take $\lambda$ to encode a total physical state of a %%@
region, then we can still state locality conditions exactly as we did originally, in Section %%@
\ref{Bellreview}. (This was Option (a) in Section \ref{formloc}; for example, cf. eq. %%@
\ref{pi2}, \ref{oi2}.) These conditions lead to a Bell inequality, ``despite our education in %%@
Budapest''.   This illustrates the present case for a common common cause, since one way to %%@
conceive such a cause (i.e. to define $\lambda$) is as the total physical state of an %%@
appropriate region.

I announced that there was also bad news. Namely, this case for a common common cause is %%@
obviously not conclusive. For my purposes, it will be enough to formulate the issue while %%@
taking the common common cause to be the total physical state of the common past; (cf. PPSI %%@
in Section \ref{prev}). Recall  that this means assuming that, once we divide the total %%@
physical state of the common past into events that do, and those that do not, causally affect %%@
the outcomes: conditionalizing on the latter does not disrupt the stochastic independence %%@
induced by conditionalizing on the former. (This was comment (1) at the end of Section %%@
\ref{prev}.) But it seems that  which events within the common past of the measurement events %%@
are causally relevant to the outcomes {\em could}  vary with the  measurement choice. And it %%@
(perhaps!) seems that this could be so, without there being any ``conspiracy'' or backwards %%@
causation. And if that is right, then which features of the total state of the common past %%@
(i.e. which partitions of the space of such states) define screeners-off---i.e. which common %%@
cause variables exist---would no doubt vary with the measurement choice: so that there is no %%@
common common cause.

 This issue takes us back again to the discussion at the end of Section \ref{formloc}: %%@
specifically, to the logical independence (mutual non-implication) of the ``fine-grained'' %%@
and ``coarse-grained'' formulations of parameter independence and outcome independence. %%@
Namely: if one endorses the previous paragraph's `seems' statements, then one will endorse %%@
the coarse-grained formulations of these conditions, which allowed different common causes %%@
for different measurement choices (i.e. eq. \ref{picoarse} and \ref{oicoarse}). And because %%@
these formulations do not imply the original fine-grained formulations (eq. \ref{pi2}, %%@
\ref{oi2}, repeating eq. \ref{pi}, \ref{oi}), it seems, on the Budapest school's evidence in %%@
Sections \ref{need} and \ref{sz}, that one will not be committed to a Bell inequality. But %%@
not all is as it seems: cf. Section \ref{avoid}.

\subsubsection{Bell inequalities from a weak PCC: the Bern school}\label{avoid}
The Budapest school's survey of known proofs and Szab\'{o}'s model (Sections \ref{need} and %%@
\ref{sz}) do not conclusively establish that {\em any} Bell's theorem needs a common common %%@
cause. Perhaps there are some plausible assumptions, not involving a strong PCC, that imply a  %%@
Bell inequality. Recent work---two papers by Grasshoff, Portmann and W\"{u}thrich, and one by %%@
Hofer-Szab\'{o}---shows that indeed there are.

 In brief, the situation is as follows. Grasshoff et al. (2005) proved a Bell inequality %%@
(viz. the Bell-Wigner inequality) from a weak PCC, other locality and no-conspiracy %%@
assumtpions, and an assumption of perfect anti-correlation for parallel settings. %%@
Hofer-Szab\'{o} (2007) develops this line, both ``negatively'' and ``positively''. %%@
Negatively, he shows that, mainly because of the assumption of perfect anti-correlation, %%@
these assumptions {\em entail} a strong PCC. (He constructs a common common cause by %%@
conjoining Grasshoff et al.'s separate common causes: cf. his equation (45).) But positively, %%@
he also shows, by a continuity argument, how to derive an analogue of the Bell-Wigner %%@
inequality from a weak PCC, without perfect anti-correlation. (The analogue adds a %%@
``correction term'' to the Bell-Wigner inequality; equation (61).) Finally, Portmann and %%@
W\"{u}thrich (2007) prove a theorem similar to Hofer-Szab\'{o}'s. But broadly speaking, their %%@
theorem is stronger in that their proof is constructive, rather than a continuity argument; %%@
and their derived inequality is an analogue of the Clauser-Horne inequality, again with %%@
correction terms (their equation (52)).

I will advertise this line of work by giving a few more details about the first paper. %%@
Grasshoff et al. assume  the weak PCC, eq. \ref{screenweak} (their PCC, p. 669). They also %%@
assume there is no correlation between conjunctions of common causes and measurement choices: %%@
that is unsurprising, since they obviously need some such assumption to block the %%@
availability of Szab\'{o}'s model. But the main assumption that enables them to overcome the %%@
logical weakness of allowing different common causes is perfect anti-correlation: if $b_1$ is %%@
parallel to $a_1$, then $pr(a_1 = +1 / b_1 = -1) = 1$ etc. They  also of course make locality %%@
assumptions. But I will not go in to detail about these, since their formulations are rather %%@
different from those of Section \ref{Bellreview}, and from Section \ref{formloc}'s %%@
coarse-grained versions (i.e. using partitions $\Lambda^m$). The difference is not just that %%@
Grasshoff et al. use the big space approach; they also formulate the assumptions in the %%@
context of their related work on (a regularity theory of) causation. So it is a good question %%@
how to rewrite their theorem in terms of Section \ref{formloc}'s conditions, or analogues: %%@
but a non-trivial question, and so not for this paper!\footnote{Suffice it to say that: (i) %%@
their theorem turns on an analogue (their Result 2, p. 673) of the standard argument by which %%@
perfect anti-correlation reduces a factorizable stochastic model to a deterministic model %%@
(e.g. Redhead 1987, pp. 101-102); and (ii) this standard argument carries over to Section %%@
\ref{formloc}'s coarse-grained versions of factorizability etc.---each choice $m = <x, y>$ %%@
with $x$ parallel to $y$ defines a partition $\Lambda^{<x,y>}$ each of whose cells is %%@
deterministic, i.e. conditioning $pr_x$ and $pr_y$ on any such cell makes probabilities %%@
trivial (0 or 1).}

Grasshoff et al. point out (2005, pp. 677)  that if perfect anti-correlation were indeed %%@
necessary for the derivation, then since it cannot be completely verified (measured %%@
correlations are never perfect), one could presumably build a Szab\'{o}-esque model %%@
incorporating a small deviation from perfect anti-correlation---which would apparently be %%@
impossible to refute in the laboratory.

But it turns out not to be necessary, as shown by the two later papers just cited. (And even %%@
if it were, it is a very reasonable assumption: for though not completely verifiable, its %%@
theoretical warrant is very strong---it arises from the conservation of angular momentum.) So %%@
the question is, as usual: which assumption of the derivation is the culprit? Though these %%@
authors do not say so explicitly, their discussions suggest they believe the culprit is the %%@
weak PCC, eq. \ref{screenweak}. I of course concur; (cf. Section \ref{Bellreview}'s %%@
discussion setting aside interpretations and loopholes). And although I have ducked out of %%@
rewriting their theorems in terms of Section \ref{formloc}'s conditions, I wager that in such %%@
a re-writing the culprit will be our coarse-grained outcome independence, eq. \ref{oicoarse}. %%@
So: at least for this paper's purposes, I conclude from these theorems, that the experimental %%@
violation of the Bell inequalities impugns the weak PCC, and more specifically, %%@
coarse-grained outcome independence. Section \ref{SELBE} will develop the consequences of %%@
this situation for SEL.

To briefly summarize this long Subsection: I have endorsed the cases made by Placek and the %%@
Bern school, that outcome dependence does impugn the PCC. Placek  targeted the strong PCC; %%@
and the Bern school  targeted the weak PCC. Now I turn to carrying over these arguments to %%@
SEL.

\subsection{SEL in the Bell experiment}\label{SELBE}
As always when one relates philosophers' conditions like PCC and SEL to an experiment, or its %%@
technical physical description, one has to exercise judgment about (i) how rigorous to be, %%@
and (ii) how well the philosophical and physical concepts mesh.

As to (i), this paper has obviously not aimed for rigour. In particular, I ducked out of the %%@
natural project in connection with the Pittsburgh-Krakow school, viz. formalizing and %%@
assessing SEL in their framework, since I think it would make no difference to my conclusions %%@
(cf. footnote 16). Sufficient unto the day is the trouble thereof!

As to (ii), we have already seen various such judgments. For example, at the end of Section %%@
\ref{Bellreview} we discussed assumptions such as taking outcomes to be distinct events, and %%@
denying the detector efficiency and locality loopholes. Only by endorsing these can we get %%@
the verdict that the Bell experiment proves outcome dependence. Judgments were also involved %%@
in taking outcome dependence to show that PCC (or PPSI or Bell's local causality) is refuted: %%@
primarily in Sections \ref{prev}, \ref{two}, \ref{avoid} and \ref{ccc}.

 I submit that it is clear enough that these discussions carry over to SEL; so that it would %%@
be very repetitive to treat them {\em seriatim}, bringing in, again {\em seriatim},  the %%@
three formulations from Section \ref{formul}, especially the two formulations of SELD, and %%@
the comparions with PCC and PPSI from Sections \ref{comp} and \ref{prev}.

Instead, I propose just to summarize  SEL's violation in the Bell experiment, under three %%@
points. The first two return us to the work reviewed in Section \ref{prev}, and so ignore the %%@
Budapest school's distinction between the weak and strong PCC.  That will prepare us for the %%@
third point, urging that this distinction does not affect the verdict that SEL is violated.

\paragraph{PCC and SEL are connected by PPSI}\label{3301} The details of my view, summarized %%@
in Section \ref{prev}, that outcome dependence refutes PCC, went hand-in-hand with outcome %%@
dependence refuting SEL. For I argued that outcome dependence refutes a cousin of PCC, dubbed %%@
PPSI. Like Bell's local causality, this took total physical states of spacetime regions %%@
(histories) as screeners-off; (it adopted the second option, `Total', of Section \ref{two}'s %%@
(Only-Total) distinction). Using total physical states, and choosing regions appropriately, %%@
meant that:\\
\indent (i) PPSI was satisfied by everyday counterexamples to PCC;\\
\indent (ii) PPSI justified parameter independence and outcome independence; and \\
\indent (iii)  the other assumptions of Bell's theorem were plausible.

The overall result was that PPSI's violation was a natural ``diagnosis'' for the %%@
``spookiness'' of the Bell correlations. The formulation  of PPSI, and the views (i)-(iii), %%@
underpinned my (1994)'s arguments that the Bell experiment violated its formulations of SEL, %%@
viz. Section \ref{formul}'s SELS and SELD2.

To this verdict, the present paper adds two points. The first, small, one is that Section %%@
\ref{formul}'s SELD1 was not formulated in my 1994. But since it implies SELD2 (Section %%@
\ref{D1impliesD2}), the violation of SELD2 implies that of SELD1. The second, main, point is %%@
that this verdict is unaffected by our education in Budapest---cf. Section \ref{3303} below.

\paragraph{The need for other judgments}\label{3302}  But I should mention that this verdict %%@
required some judgments that in this paper I have so far not mentioned, since they do not %%@
relate to the issue of common common causes. These are judgments that I still endorse but %%@
which, I agree, can be denied. Though I need not repeat the details, I should signal that %%@
they are of two kinds. First, there is a judgment only about SEL; then there are judgments %%@
about probabilities, which  concern PCC and SEL equally.

\indent (i): One can formulate the broad idea of SEL (at the start of Section \ref{selintro}) %%@
differently from Section \ref{3formul}'s three versions; and some such formulations are {\em %%@
not} violated by the Bell experiment. Indeed, Hellman's original paper had formulations of %%@
SEL analogous to our SELS and SELD2 (his (4) and (5); 1982, p. 466, 495-497). But (as I %%@
mentioned in footnote 5) he wanted to deny that the Bell experiment involved spacelike %%@
causation, or  any  other kind of ``spookiness''. So he hedged his formulations of SEL with %%@
provisos, so that they were not violated by the experiment; (nor by similar unscreenable-off %%@
correlations, for example in a toy-example of Hellman's about a ``flashing'' particle). So %%@
deploying my formulations of SEL, rather than Hellman's with his provisos, involves a %%@
judgment: the outcome dependence in the Bell experiment, and Hellman's flashing particle %%@
example, are ``spooky''. More precisely, they are spooky enough that one should prefer a %%@
formulation of SEL that they violate. (For details of this judgment, and the provisos and %%@
example, cf. my (1994, Sections 6 and 7).)

\indent (ii): Finally, my previous verdicts, both for PCC and SEL, needed some judgments %%@
comparing empirical and theoretical probabilities. These judgments are uncontroversial, and %%@
in no way special to my position: but they are worth articulating just because many %%@
discussions are silent about them. The sort of judgment needed is shown by [a] and [b].\\
\indent \indent [a]: The trio $<\lambda, x, y>$ of hidden variable and choices is: for PCC, %%@
an appropriate common cause; or for SEL: an appropriate history $H$; (and similarly for local %%@
causality). Or more exactly: though $<\lambda, x, y>$ may not itself be the appropriate %%@
common cause or history, a probability function determined by $<\lambda, x, y>$ (viz. by some %%@
sort of averaging over other factors that should also contribute to the common cause or %%@
history) gives probabilities for outcomes that are close to those that would be given by PCC %%@
or SEL. Here, `close' means of course `sufficiently close' , as determined by [b]... \\
\indent \indent [b]:  The probabilities given by SEL (or PCC, or local causality) differ %%@
enough from the quantum (or better: experimentally right!) probabilities that  outcome %%@
dependence in the experiment refutes the screening-off equation of, say, SEL. More exactly:  %%@
outcome dependence refutes, in the first place, the screening-off equation with probabilities %%@
determined by $<\lambda, x, y>$; then by [a], it refutes the screening-off equation of  SEL %%@
or PCC or Bell's local causality.

To sum up these two Paragraphs: I agree that taking the Bell experiment to violate SEL %%@
requires questionable judgments---which I have tried to articulate, here and especially in my %%@
(1994).

\paragraph{Weak vs. strong SELD}\label{3303} How if at all does our education in %%@
Budapest---i.e. the distinction between weak and strong PCC, or the question whether there is %%@
a common common cause---affect this verdict? I submit that it does {\em not}.

The reason lies in two points from Section \ref{B}.  First: both PPSI and SEL sidestep the %%@
distinction by having a probability distribution prescribed by an individual world, or %%@
history, or region's total physical state. We saw this for SEL in Section \ref{formul}, for %%@
PPSI in Comment (2b) at the end of Section \ref{prev}, and for locality conditions tailored %%@
to the Bell experiment in Option (a) of Section \ref{formloc}. Second: this feature of PPSI, %%@
SEL and Option (a) was defended in Section \ref{ccc}, following Placek; (though with the %%@
concession that a  Simpson's paradox threat might prompt resort to the Bern school's %%@
theorems: Section \ref{avoid}).

Of course, this is not to deny that one {\em could} introduce partitions $\Lambda$ of the %%@
probability space into the formulation of PPSI and SEL, and so write down weak and strong %%@
versions of them. And agreed, if one does so: (i) the earlier formulations will correspond to %%@
strong versions (with the partition given by the singleton sets of each $\lambda$); (ii) my %%@
verdict that PPSI and SEL are violated will hold good only for strong versions of %%@
SEL---again, as we would expect from Section \ref{B}, especially Szab\'{o}'s model (Section %%@
\ref{sz}).

For the sake of completeness, I should exhibit such strong formulations. For simplicity, I %%@
will again consider only the traditional Bell experiment with four correlated pairs $(X,Y) = %%@
(A_1,B_1),...,(A_2,B_2)$ (Section \ref{Bellreview}). So I will not consider SELS. We envisage %%@
that the possible  L-outcomes $A_1, A_2$ occur in the same spacetime region, and similarly on %%@
the R-wing, so that we can specify a region as `the common past of $X$ and $Y$'. That is: %%@
there is for PPSI the usual ambiguity about what to mean by `common past'---what if anything %%@
beyond $C^-(X) \cap C^-(Y)$ to include. But there is no {\em further} ambiguity in the %%@
definitions of $C^-(X)$ and $C^-(Y)$.

 I shall also assume that for PPSI as well as SEL, we are to use partitions of the set $\cal %%@
W$ of worlds to encode the idea that only some features of the common past of $X$ and $Y$, or %%@
of history up to a hypersurface $t$, or of history within a ``Table Mountain'' region $C^-(t) %%@
\cap C^-(X)$, determine the probabilities of $X$ and $Y$ and their Boolean combinations. That %%@
is: worlds in the same cell of the partition match as regards these features, and so match as %%@
regards these probabilities. (We are of course back again at the idea that there is no %%@
`Simpson event', conditionalizing on which would break the stochastic independence got by %%@
conditionalizing on a cell of the partition.) But I will not make my formulations of PPSI and %%@
SEL express explicitly the idea that the partition encodes features confined to a part of %%@
spacetime (viz. the common past of $X$ and $Y$, or the past of a hypersurface $t$, or a %%@
``Table Mountain'' region $C^-(t) \cap C^-(X)$): this would of course be a matter of %%@
demanding that within each cell of the partition, any possible history of the  {\em rest} of %%@
spacetime is included in some element of the cell.\footnote{As mentioned in footnote 16, the %%@
Pittsburgh-Krakow framework makes rigorous sense of `same spacetime region' etc., and so %%@
gives the materials for formulations of PPSI and SEL that express explicitly the partition %%@
being about only a part of spacetime. But this paper ducks out of developing the details.}

With these assumptions, a strong version of PPSI would be as follows:
\begin{quote}
{\bf Strong PPSI}: There is a partition $\Lambda$ of the set $\cal W$ of worlds comprising a %%@
Bell experiment, with each cell $\lambda \in \Lambda$ encoding features of the common past of %%@
the outcomes $X = A_1, A_2$ and $Y = B_1, B_2$, and such that for all outcomes $X$ and $Y$:
\be
pr(X \& Y / \lambda) = pr(X / \lambda) \cdot pr(Y / \lambda ) \; .
\label{strongppsi}
\ee
\end{quote}
In effect, we have just combined Section \ref{prev}'s PPSI, eq. \ref{ppsi}, with Section %%@
\ref{resus}'s strong PCC, eq. \ref{screenstrong}.

Similarly for a strong version of SEL; with the difference that we can preserve Section %%@
\ref{formul}'s explicit indexing of probability distributions with either times or histories %%@
(in addition to worlds), and its universal quantification over these indices. This universal %%@
quantifier `for all times $t$/histories $H$', must of course come before the existential %%@
quantifier, that there is a partition: which must itself come before the universal quantifier %%@
`for all outcomes', in order that the formulation be strong. Thus we get for SELD1 and SELD2:

\begin{quote}
{\bf Strong SELD1}: Given the set $\cal W$ of worlds comprising a Bell experiment with both %%@
outcomes $X$ (respectively $Y$) occurring in the same spacetime region ($X$ and $Y$ %%@
spacelike):\\
for any hypersurface $t$ earlier than both $X$ and $Y$, and dividing $C^-(X)$, and  such that %%@
$C^-(X) \cap C^-(Y) \cap C^+(t) = \emptyset$:\\
there is a partition $\Lambda$ of  $\cal W$, with each cell $\lambda \in \Lambda$ encoding %%@
features of the  past of $t$, such that for all outcomes $X$ and $Y$
\be
pr_{t}(X \& Y / \lambda) = pr_{t}(X / \lambda) \cdot pr_{t}(Y / \lambda ) \; .
\label{seld1str}
\ee
\end{quote}

\begin{quote}
{\bf Strong SELD2}:  Given the set $\cal W$ of worlds comprising a Bell experiment with both %%@
outcomes $X$ (respectively $Y$) occurring in the same spacetime region ($X$ and $Y$ %%@
spacelike):\\
for any hypersurface $t$ earlier than $X$, and dividing $C^-(X)$, and such that $Y$ is in the %%@
difference $C^-(t) - C^-(X)$:\\
there is a partition $\Lambda$ of  $\cal W$, with each cell $\lambda \in \Lambda$ encoding %%@
features of the history in the intersection $C^-(t) \cap C^-(X)$, such that for all outcomes %%@
$X$ and $Y$
\be
pr_{t}(X \& Y / \lambda) = pr_{t}(X / \lambda) \cdot pr_{t}(Y / \lambda ) \; .
\label{seld2str}
\ee
\end{quote}

Note that though there is a time index in eq.s \ref{seld1str} and \ref{seld2str}, it is %%@
``formal'' in the sense that $pr_t$ cannot be interpreted as it was in Section \ref{formul}'s %%@
formulations: viz. as a probability distribution that incorporates  all the ways that  %%@
history up to $t$ bears on probabilities of events future to $t$. For if $pr_t$ incorporated %%@
all that, then the conditionalization on $\lambda$ in eq.s \ref{seld1str} and \ref{seld2str} %%@
would not be conditionalization on new information, that was not already incorporated in %%@
$pr_t$---while the idea of strong SELD1 and SELD2 is that $\lambda$ should represent new %%@
information.\footnote{Indeed, if probabilities evolve by conditionalization on intervening %%@
history, eq. \ref{evolvecond}, then this interpretation of $pr_t$ would mean that any %%@
$\lambda$ is already either implied or excluded by $pr_t$, i.e. $pr_t(X/\lambda) = pr_t(X)$ %%@
etc., or $pr_t(\lambda) = 0$.}

With these formulations in hand, we can ask about their relations, as we did in Section %%@
\ref{314rel.A}. In considering this, we have to bear in  mind that since the time index is %%@
now ``formal'', assumptions that in Section \ref{314rel.A} were compelling or plausible may %%@
well now not be. But for one implication, we are ``lucky''. It is easy to check that Section %%@
\ref{D1impliesD2}'s proof that SELD1 implies SELD2 carries over intact to the strong %%@
versions; (for it depends on relations between hypersurfaces, not at all on the %%@
interpretation of $pr_t$). That is: Strong SELD1 implies Strong SELD2. On the other hand, %%@
Section \ref{D2impliesD1}'s proof that SELD2 implies SELD1, given two assumptions, falls %%@
down. For both assumptions depend on the interpretation of $pr_t$; (the first was that %%@
probabilities evolve by conditionalization on intervening history). Maybe there are plausible %%@
analogous assumptions that would give a proof.

But I shall not pursue details. Space presses, and I want to close by glimpsing how SEL fares %%@
beyond quantum mechanics: in algebraic quantum field theory (Section \ref{causeaqft}) and in %%@
dynamical spacetimes (Section \ref{causet}). My aim will be very modest: to advertise some %%@
recent work and open problems; and to save space, I will make no attempt to explain the %%@
relevant frameworks and formalisms.

\section{SEL in Algebraic Quantum Field Theory}\label{causeaqft}
\subsection{The story so far}\label{sofar2}
As I mentioned in Section \ref{Intr}, AQFT provides a more precise conception of events, %%@
causal influences and probabilities than the philosophical conception used in Sections %%@
\ref{sel} and \ref{qmphys}. For us, this has two main consequences. The first is a  general %%@
point about the broad idea of stochastic relativistic causality. Namely, one can write down %%@
several obviously different formulations of the idea, and go on to prove them to be logically %%@
independent of one another (even in the presence of other axioms of the framework). Besides, %%@
this sort of independence also holds good for corresponding formulations in heuristic quantum %%@
field theory. (For some details, cf. Horuzhy (1990, pp. 19-21); Butterfield (2007, Section 3) %%@
is a philosopher's introduction.) So I stress that, even setting aside our interest in %%@
formulations of SEL, the quantum field theory literature yields rich pickings about %%@
stochastic relativistic causality.

The second consequence is specific to our topic, SEL. Because the conception of events etc. %%@
is more precise, fewer judgments are required about how best to ``transcribe'' SEL into AQFT %%@
than into quantum mechanics. In particular, it is pretty clear how best to transcribe Section %%@
\ref{formul}'s formulations of SEL into AQFT; and it is clear which formulations are obeyed %%@
and which violated.

This was the topic of my (1996). To summarize: I argued that the most natural transcription %%@
of SELS is provably {\em satisfied}.  But the natural transcription of SELD2 is violated---on %%@
account of the outcome dependence (violations of Bell inequalities) exhibited (endemically) %%@
by quantum field quantities, shown in various papers by Landau, Summers, Werner et %%@
al.\footnote{The points about SELS and SELD were prompted by a disagreement with previous %%@
literature; cf. R\'{e}dei 1991 and Muller and Butterfield (1994). As I mentioned in Section %%@
\ref{formul}, there is a mnemonic: `S' in SELS stands for `satisfied' as well as `single', %%@
and `D' in SELD stands for `denied' as well as `double'. For a full survey of the results of %%@
Landau, Summers, Werner et al. upto ca. 1990, cf. Summers (1990). Later work includes %%@
Halvorson and Clifton (2000).} I need not quote these transcriptions. For our purposes, it is %%@
enough to say the following:\\
\indent (i): The worlds of SELS and SELD are replaced by models of an algebraic quantum field %%@
theory, with each model given by a triple $< {\cal M}, {\cal A}, \phi >$ where $\cal M$ is %%@
the spacetime (in my 1996: always Minkowski spacetime), $\cal A$ is the assignment to each %%@
open bounded region $O$ of $\cal M$ of a local algebra ${\cal A}(O)$, and $\phi$ is the state %%@
(expectation functional). Accordingly, an event $E$ is replaced by a projector in a local %%@
algebra, and $E$'s probability by  the projector's expectation in $\phi$. We think of the %%@
models as satisfying (making true) various sentences of a putative formal language expressing %%@
AQFT; and among these sentences will be ascriptions of expectation values of local %%@
projectors.\\
\indent (ii): How should we transcribe the idea of the probability prescribed by history up %%@
to $t$, or by history within the ``Table Mountain'' $C^-(E) \cap C^-(t)$, and similar ideas? %%@
The most obvious tactic, viz. conditionalizing $\phi$ on many projectors associated with the %%@
region, is clearly fraught with difficulties, both mathematical and interpretative (i.e. %%@
about value ascriptions in quantum theory): for example, are we to conditionalize $\phi$ on %%@
one of every pair of a projector and its orthocomplement?\\
\indent (iii): But fortunately, we can avoid these difficulties by using the idea that models %%@
match in their history in a spacetime region if they make true the same set of sentences %%@
about the region. That is, we formulate SEL as a universally quantified conditional, along %%@
the lines: `For any two models on the same spacetime $< {\cal M}, {\cal A}_1, \phi_2 >$, $< %%@
{\cal M}, {\cal A}_2, \phi_2 >$, for any region $O \subset {\cal M}$, and any projector $E %%@
\in {\cal A}(O)$: if the two models match throughout a suitable region $O'$ whose future %%@
domain of dependence  $D^+(O')$ includes $O$, $O \subset D^+(O')$, then ...'. Thus the %%@
matching on the region $O'$ expresses the restriction to probabilities conditional on %%@
sufficiently rich information about the past.\\
\indent (iv): The tactic in (iii) is common to the transcriptions of SELS and SELD2. The main %%@
difference between the transcriptions is then that SELS is about just the expectation of $E %%@
\in {\cal A}(O)$, so that SELS's consequent states that the two models match in their %%@
expectations for $E$, i.e. $\phi_1(E) = \phi_2(E)$;  while
SELD2 is about the equality of $E$'s expectation, with its expectation conditional on another %%@
projector, $F$ say, associated with a suitable region spacelike to $O$, so that SELD2's %%@
consequent states that in each model $\phi_j(E) = \phi_j(E/F), j = 1,2$. (We need not discuss %%@
other minor differences, and the definition of `suitable region'.)\\
\indent (v): As mentioned, the main results are that, thus transcribed, SELS follows directly %%@
from two axioms of AQFT (the Isotony and Diamond axioms); and that SELD2 is endemically %%@
violated, thanks to the endemic Bell correlations (outcome dependence) shown by Landau, %%@
Summers, Werner et al.

This verdict meshes happily with that of Section \ref{SELBE}: Bell correlations (outcome %%@
dependence) violate SEL in a ``double'' formulation, in field theory, just as in quantum %%@
mechanics; again providing an appropriate diagnosis of the strangeness of the Bell %%@
correlations. Besides, field theory's more precise framework makes the verdict more secure, %%@
i.e. less dependent on extraneous judgments.

\subsection{Questions}\label{qns}
But as in Section \ref{B}, this verdict needs to be reviewed in the light of the subsequent %%@
literature. I will briefly discuss this, focussing only on questions raised by: the %%@
formulations in Section \ref{sel} (Section \ref{ourfmns}); the work of the Budapest and Bern %%@
schools (Section \ref{thebs}). In both areas, it will be clear that there are open questions.

\subsubsection{Our formulations}\label{ourfmns}
Section \ref{sel} presented formulations of SEL that applied not only to Minkowski spacetime %%@
but to any stably causal, in particular globally hyperbolic, spacetime. So the main technical %%@
question raised by considering SEL for AQFT is whether the results of Landau et al. about %%@
endemic outcome dependence can be generalized from Minkowski spacetime to other spacetimes. %%@
This is a large vague question, and I shall not pursue it; but the short answer is that they %%@
can be. A bit more precisely:  an endemic violation of Bell inequalities (and so, {\em %%@
modulo} our usual assumptions: endemic outcome dependence) can be shown in models using %%@
globally hyperbolic spacetimes. For example, cf. Proposition 4 of Halvorson and Clifton %%@
(2000). (For an introduction to the algebraic formulation of quantum field theory on curved %%@
spacetimes, and its advantages, cf. Wald (1994: pp. 53-65, 73-85).)

 Section \ref{sel} raises two other obvious questions about my (1996)'s discussion. First, we %%@
can ask about Section \ref{formul}'s  SELD1, whose transcription to AQFT was not discussed in %%@
(1996); (since SELD1 was not formulated in my 1994). Again, I will not pursue details since %%@
the situation is straightforward. One can transcribe SELD1 into AQFT along the same %%@
lines---and as naturally---as my (1996) transcribed SELD2; and again, endemic outcome %%@
dependence shows that SELD1 is violated.

Besides, we saw in Section \ref{D1impliesD2} that SELD1 implies SELD2 by a short argument %%@
which turned just on spatiotemporal relations between the events $E$ and $F$ and the %%@
hypersurface $t$. Though I will not give details, it is straightforward to show that SELD1 %%@
and SELD2 can be transcribed into AQFT in such a way that this argument carries over. So %%@
using these formulations, AQFT's violation of (the transcription of) SELD1 will follow from %%@
its violation of (the transcription of) SELD2, by {\em modus tollens}.

 The second obvious question raised by Section \ref{sel} concerns Section \ref{314rel.B}'s %%@
proof of equivalence of SELS and SELD2. Since the former holds, and the latter fails, once %%@
they are transcribed in AQFT, we know this proof must break down in AQFT. And there is no %%@
mystery. Both its main assumptions, viz.\\
\indent (i) probabilities evolve by conditionalization eq. \ref{evolvecond};\\
\indent (ii) all the worlds in $\cal W$ have the same initial probability function $pr$; \\
evidently fail in AQFT: i.e. once transcribed in the obvious way suggested by (i)-(iii) in %%@
Section \ref{sofar2}.\footnote{Besides, the proof's two simplifying assumptions, that there %%@
are finitely many possible histories ((a) and (b) of Section \ref {314rel.B}) may carry over %%@
less well to AQFT, than to realistic classical theories. But even if there are no such %%@
cardinality issues, the conceptual assumptions (i) and (ii) fail.}

\subsubsection{The Budapest and Bern schools}\label{thebs}
I turn to three questions raised by the work of the Budapest and Bern schools. The first is %%@
obvious in the light of the discussion of Section \ref{B} and \ref{SELBE}. There we learnt:\\
\indent (i): from the Budapest school that outcome dependence impugns primarily the strong %%@
PCC not the weak one; but also;\\
\indent (ii): one could reply that the strong PCC was plausible (Section \ref{ccc}), and that %%@
SEL reflected this by having a probability distribution prescribed by an individual world, or %%@
history, or region's total physical state (Section \ref{3303} and previous discussions cited %%@
there).

\indent So now the obvious question is: do (i) and (ii) carry over to AQFT? I think it is %%@
clear that Yes, they do; though again I will not enter into details. That is:\\
\indent (i'): A Bell's theorem for field-theoretic quantities pertaining to spacelike regions %%@
would normally proceed from a strong PCC. (For details, cf. eq. 3.6 - 3.8 in my (1996).) %%@
Here, I say `normally', since you might ask: what about adapting the Bern school's theorems %%@
(Section \ref{avoid}) to AQFT? For this, cf. question (2) below.)\\
\indent (ii'): But again, a strong PCC is plausible; and once SELD is transcribed to AQFT in %%@
terms of two models matching in all the sentences about a region that they make true ((iii) %%@
in Section \ref{sofar2}), it reflects the strong PCC in the same way as we saw in Section %%@
\ref{3303}.

The second and third questions arise from a result of R\'{e}dei and Summers (2002), showing %%@
that in a certain sense the weak PCC is provably satisfied, {\em not} violated, by AQFT. More %%@
precisely, they show that: for any AQFT, i.e. assignment of local algebras to regions $O %%@
\subset {\cal M} \mapsto {\cal A}(O)$, subject to standard conditions; for any spacetime %%@
regions $O_1, O_2$ contained in a pair of spacelike double cones;  for any state $\phi$ of a %%@
standard type (viz. locally normal and faithful); for any pair of projectors $E \in {\cal %%@
A}(O_1), F \in {\cal A}(O_2)$: if $E$ and $F$ are correlated, i.e.
\be
\label{RScrn}
\phi(E \wedge F) > \phi(E)\phi(F)
\ee
then there is a projector $C$ in (the algebra for) the weak causal past of $O_1$ and $O_2$, %%@
i.e. the region $(C^-(O_1) - O_1) \; \cup \; (C^-(O_2) - O_2)$, consisting of points which %%@
can causally influence some point in either $O_1$ or $O_2$, that screens off the correlation %%@
in eq. \ref{RScrn}. That is:
\be
\label{RSscrnoff}
\phi(E \wedge F \; / \; C ) = \phi(E\; / \; C )\phi(F\; / \; C ) \; .
\ee

This prompts two questions:\\
\indent (1): How can this result be reconciled  with the gist of our previous discussion, %%@
that SELD is violated? And:\\
\indent (2): How can this result be reconciled  with Hofer-Szab\'{o}'s and the Bern school's %%@
theorems (Section \ref{avoid}), that a Bell's theorem can be proved using a weak PCC? (There %%@
is a tension here since AQFT endemically violates Bell inequalities, so that we expect it to %%@
violate some assumption or other of any Bell's theorem.)\\
\indent I shall briefly state my view about the answers to these questions; but again, it %%@
will be obvious that there are open questions hereabouts.

(1): I think the reconciliation lies in two contrasts between Redei and Summers' result, and %%@
the violation of SELD (as transcribed to AQFT). Both are obvious. The first relates to the %%@
distinction between the weak and strong PCC, which dominated Section \ref{qmphys}. As noted %%@
in (ii') just above, SELD reflects a strong PCC. But R\'{e}dei and Summers prove a weak PCC: %%@
for each $E, F$ pair, there is a screener-off $C$.\footnote{They of course recognize this, %%@
and see it as developing the Budapest school's critique of the Bell literature's conflating %%@
the weak/strong distinction. In jocular terms: they would be worried if they could prove a %%@
strong PCC, for fear of a Bell's theorem.}

Second, the two results use very different ideas of ``conditionalizing on the screener-off''. %%@
R\'{e}dei and Summers conditionalize in the usual sense (albeit adapted to quantum states) on %%@
an event (projector) $C$. But ever since Section \ref{selintro}, the idea of SEL, like that %%@
of Bell's local causality, is to consider the probability prescribed by the whole history of %%@
a (large) region. And as reported in (ii) and (iii) of Section \ref{sofar}, this is %%@
transcribed into AQFT, not in terms of the (problematic) idea of conditioning on some sort of %%@
``complete'' set of projectors for the region, but in terms of matching of models of the %%@
theory.

A side-remark. Beware: reading R\'{e}dei and Summers' theorem, you might think that the %%@
reconciliation lies elsewhere---in a contrast about the spatiotemporal location of the %%@
screener-off. For R\'{e}dei and Summers' screener-off $C$ is associated with the weak past of %%@
the regions $O_1$ and $O_2$. But SELD (transcribed to AQFT) requires matching of models on a %%@
region much smaller than the weak past. (In fact, in (iii) of Section \ref{sofar}, I talked %%@
vaguely about `a suitable region' $O'$, saying only that its future domain of dependence %%@
$D^+(O')$ must satisfy a certain condition: but the details in my (1996) make it clear that %%@
the region of required matching is indeed much smaller than the weak past.) But beware: this %%@
contrast is not the reason (or even: ``a third reason'') for   the reconciliation. For as my %%@
1996 brings out, in AQFT's violation of SELD the region of matching can be taken to be much %%@
larger than the weak past. In other words: AQFT also violates a version of SELD that is %%@
apparently weaker  than my transcription, since it assumes models' matching on a much larger %%@
region. Accordingly, the reason for the reconciliation must be sought elsewhere: by my %%@
lights, in the two contrasts above.

(2): I turn to the question why Section \ref{avoid}'s Bell's theorems, assuming a weak PCC, %%@
fail to carry over to AQFT. (As I presume they must, on pain of their conjunction with %%@
R\'{e}dei and Summers' theorem contradicting the endemic violation of Bell inequalities.) %%@
Here I am less confident than for question (1), and will just endorse a suggestion of %%@
Portmann's (to whom my thanks). Namely: the assumption that conjunctions of common causes %%@
(for different correlations) are statistically independent of measurement choices {\em fails} %%@
in AQFT, once transcribed in the natural way with common cause projectors $C$ in the sense of %%@
R\'{e}dei and Summers. (This is assumption 5 in Portmann and W\"{u}thrich (2007), and %%@
equation (58) in Hofer-Szab\'{o} (2007). Recall that Szab\'{o}'s model shows it is needed for %%@
a Bell's theorem.) Indeed, we would expect it to fail in so far as two common cause %%@
projectors $C_1$ and $C_2$ will in general not commute, and so will lack a joint quantum %%@
probability distribution.\footnote{For the Bern school's earlier theorem (Grasshoff et al. %%@
2005) which assumed perfect anti-correlation, there may also be a reconciliation in the %%@
failure of {\em that} assumption in AQFT. For so far as I know, the best we can do for this %%@
assumption is along the lines that for any $\epsilon > 0$, there is a pair of %%@
(spacelike-related) projectors which are ``1 - $\epsilon$'' correlated; (Redhead, 1995,  %%@
Theorem 4', p. 133).}

\section{SEL in dynamical spacetimes}\label{causet}
Among the many approaches to ``quantum gravity'', i.e. the reconciliation of quantum theory %%@
and general relativity, there are some that try to combine directly the ideas of %%@
stochasticity (from quantum theory) and dynamical spacetime, i.e. the metric representing %%@
gravity and so being responsive to matter (from general relativity). In these approaches, SEL %%@
and related ideas like Bell's local causality of course play a central role.

I will end by indicating two examples. In Section \ref{kent}, I report that Kent (2005) uses %%@
a formulation of SEL (albeit under another name)  appropriate to such a spacetime, so as to %%@
propose a (doable!) experiment exploring the interaction of quantum mechanics and general %%@
relativity (or any theory that geometrizes gravity and so has a dynamical  spacetime). Then %%@
in Section \ref{causets}, I describe how for the causal set programme, developed by Sorkin %%@
and others, the naive transcription of SEL fails trivially---prompting the search for better %%@
causal-set formulations of relativistic causality.

\subsection{SEL for metric structure?}\label{kent}
Kent (2005a) does not mention SEL: he works with a formulation of Bell's local causality %%@
appropriate for theories with a dynamical metric like general relativity. But it will be %%@
clear that his discussion could be recast in terms of our SELD1 and SELD2 (cf. Section %%@
\ref{comp}), and their refutation by the Bell experiment (Section \ref{SELBE}).

Kent proposes (following a suggestion of Dowker) that Bell's condition be adapted to a %%@
stochastic geometrized theory of gravity, as follows. Let $\lambda$ be a spacetime region %%@
equipped with specified metric and matter fields, that contains its own causal past. Let %%@
$\kappa$ be any spacetime region with specified metric and matter fields. Let $pr(\kappa / %%@
\lambda)$ be the probability that the domain of dependence $D(\lambda)$ of $\lambda$ is %%@
isometric to $\kappa$. Let $\kappa '$ be any other region of spacetime with specified metric %%@
and matter fields, which we know to be spacelike from $D(\lambda)$. Then we say that the %%@
envisaged theory of spacetime is {\em locally causal} iff for all such $\lambda, \kappa, %%@
\kappa '$, we have
\be
pr(\kappa / \lambda) = pr(\kappa / \lambda \& \kappa ').
\label{kent}
\ee

Kent now makes three points.\\
\indent (1): General relativity is locally causal, thanks to the metric and matter fields in %%@
$D(\lambda)$ being completely determined by those in $\lambda$. (Of course a wealth of %%@
mathematical physics, about the well-posedness of initial value problems, lies behind this %%@
point. But we do not need details.)\\
\indent (2): But now imagine a standard Bell experiment, with the wings spacelike related, in %%@
which each detector is coupled to a nearby Cavendish experiment, so that in each wing each %%@
measurement choice and outcome leads to
\begin{quote}
one of four different configurations of lead spheres---configurations which we
know would, if the experiment were
performed in isolation, produce one of four macroscopically and testably
distinct local gravitational fields ...
The separation of the two wings is such that the gravitational field
test on either wing can be completed in a region space-like separated
from the region in which the photon on the other wing is detected ... Extrapolating any of %%@
the standard interpretations
of quantum theory to this situation, we should expect to see precisely the same
joint probabilities for the possible values of the gravitational fields in
each wing's experiments as we should for the corresponding outcomes in the
original Bell experiment ... Then, if $\kappa$ is the region immediately surrounding the %%@
measurement choice
and outcome in one wing of the experiment, $\kappa'$ the corresponding
region for the other wing, and $\Lambda$ the past of $\kappa$, we have
$$
{\rm Prob}( \kappa | \Lambda )
\neq {\rm Prob}( \kappa | \Lambda , \kappa' ) \, .
$$
... [Thus] we have a direct conflict
between the predictions of two outstandingly successful theories: quantum
theory and general relativity. (Kent 2005a, p. 2)
\end{quote}

(3): Kent then considers what are the conceivable experimental outcomes. There are several %%@
possibilities, but I shall just quote the two obvious ones---and recommend Kent's discussion %%@
for more details.
\begin{quote}
One is that the violations of local causality predicted by
quantum theory, and to be expected if some quantum theory
of gravity holds true, are indeed observed.  This would
empirically refute a key feature of general
relativity, namely, the local causality of space-time ...
 [A second conceivable outcome is] ... that the measurement results obtained
from the detectors fail to violate the [Bell] inequality.
This would imply that quantum theory fails to describe correctly
the results of the Bell experiment embedded within this particular
experimental configuration, and so would imply a definite limit
to the domain of validity of quantum theory. (Kent 2005a, p. 3).
\end{quote}

\subsection{SEL for causal sets?}\label{causets}
\subsubsection{The causal set approach}\label{causetintro}
I turn to the causal set programme.\footnote{For this Section's purposes, the main references %%@
are: Brightwell et al. (2002), Dowker (2005), and Rideout and Sorkin (2000).} It models %%@
spacetime as a discrete set of spacetime points, partially ordered by causal connectibility, %%@
which ``grows'' by a stochastic process of adding points to the future of the given discrete %%@
set. The set is called a `causal set', or for short {\em causet}. In more detail: a causet is %%@
a partially ordered set (poset) $C$ which is locally finite. The partial order represents %%@
causal connectibility; and `locally finite' means that, writing the partial order as $x \prec %%@
y$, we require that for any $x, y$, the set $\{z: x \prec z \prec y\}$ is finite. We think of %%@
this causal structure as growing in successive stages. At any given stage, a new element gets %%@
added to the immediate future of some of the given elements. Here `some' (i) means, in %%@
general, not all, and (ii) allows `none': i.e. the new element can be spacelike to all the %%@
given elements. Then at the next stage, another new element is added to (the immediate future %%@
of some of) the augmented set; and so on. Furthermore, this growth is stochastic: there are %%@
to be probabilistic rules for the various ways of adding an element. We write $\Omega(n)$ for %%@
the set of $n$-element causets, i.e. posets with $n$ elements which are {\em a fortiori} %%@
locally finite. So with $C \in {\Omega(n)}$ we envisage
 probabilistic rules for transitions: $C \in {\Omega(n)} \rightarrow C' \in {\Omega(n+1)}$.

Of course, we also envisage that matter fields should be added to this framework of %%@
stochastic  causal structure. But plenty of interesting questions, both conceptual and %%@
technical, can be formulated and attacked without adding a representation of matter to the %%@
framework. I shall present just one, viz. the specification of the probabilistic rules (the %%@
dynamics), since it relates to SEL.

Besides, even without adding matter, there are three general motivations for investigating %%@
this framework, which are worth stressing before discussing dynamics. First, there are %%@
various results to the effect that the causal structure of a general relativistic spacetime %%@
``almost determines'' its metric structure (so that philosophers' traditional idea of a %%@
``causal theory of time'' is ``almost right''). These results make it reasonable to take as a %%@
``toy-model'' of stochastic metric structure of the sort we expect quantum gravity to %%@
require, a stochastic causal structure.

 Second, the causet framework provides analogues of general relativistic ideas, such as %%@
general covariance and the definition of observables, whose bearing on quantum gravity is %%@
agreed to be important and problematic. So causets provide a ``toy-model'' or ``conceptual %%@
laboratory'' where  issues about these ideas can be investigated. For us, the  important  %%@
example of this will be that causets raise good questions about how to formulate relativistic %%@
causality for a dynamical spacetime: in particular, the naive transcription of SEL fails.

Third, a point for philosophers who (like me) endorse the ``block-universe'' or ``B-theory'' %%@
of time. I (together with my informants who work on causets!) take the causet approach to be %%@
consistent with this metaphysical view---despite its talk about `spacetime growing', %%@
`spacetime points being born' etc. Indeed, there are two points here. The first is not %%@
specific to causets: it applies equally to classical stochastic process theory, or indeed %%@
indeterminism in general, and is familiar to philosophers: viz., these are consistent with %%@
the block-universe---despite their talk about one of many future alternatives `coming to be'. %%@
The second point is that in the causet approach, the successive stages of a causet's growth %%@
are ``gauge'': they are not intended to be physically significant. This point is taken up in %%@
Section \ref{label}.

\subsubsection{Labelled Causal Sets; General Covariance}\label{label}
To define a stochastic evolution on causets, it has hitherto been indispensable to think in %%@
terms of successive stages, at each of which a new point added to the given causet. So each %%@
element/point of each causet in such a process (in the jargon of stochastic process theory: %%@
such a realization, trajectory) is labelled by a natural number $n \in {\cal N}$, %%@
representing the stage at which it was added. Writing $l(x) \in {\cal N}$ for the label of %%@
the point $x$, the labelling therefore obeys: $x \prec y \Rightarrow l(x) < l(y)$. (Of %%@
course,  the converse implication fails since the point added later, viz. $y$, might be added %%@
so as to be spacelike to $x$.) So each labelled causet:\\
\indent (i): has a ``birth'' (``big bang''), when at stage 1, the trivial poset, the %%@
singleton set of one spacetime point, is ``born''; and so:\\
\indent (ii): is finite towards the past; (but we allow the stochastic process to run to %%@
infinity, so that we can consider denumerable posets)\\
\indent (iii): determines an upward path through the poset of all finite {\em un}labelled %%@
posets (ordered by inclusion in the obvious sense): so again writing $\Omega(n)$ for the set %%@
of $n$-element unlabelled causets, we write this poset of all finite unlabelled causets as %%@
$\Omega({\cal N}) := \cup_{n \in {\cal N}}\Omega(n)$. (So all the $n$-element causets form %%@
rank $n$ of $\Omega({\cal N})$.)

The process of growth on labelled causets is to be non-deterministic, in the sense that
for each labelled causet ${\tilde C}_n$ of $n$ elements, and each subset $S \subset {\tilde %%@
C}_n$ that is closed under taking of ``ancestors''  (i.e. if $y \in S$ and $x \prec y$ then %%@
$x \in S$):
 the next stage, labelled $n+1$,  could add its new element just to the future of $S$'s %%@
maximal elements. In an obvious jargon: stage $n+1$ could choose $S$ as its newborn element's %%@
{\em precursor} set.

Furthermore, the process of growth is to be stochastic in that for all ${\tilde C}_n$ and for %%@
all such closed-under-ancestors subsets $S \subset {\tilde C}_n$, there is to be a %%@
corresponding transition probability. The set of all these transition probabilities fixes a %%@
dynamics.\footnote{To be precise: it can be shown that a specification of all
these transition probabilities fixes a stochastic process on the $\sigma$-algebra generated %%@
by the cylinder sets built on all the finite labelled causets. These measure-theoretic %%@
notions are crucial for the causet approach's characterization of observables; for this %%@
topic, cf. Brightwell et al. (2002, 2003).}

I said that to define this stochastic dynamics, it has hitherto been indispensable to think %%@
in terms of stages, i.e. to use labelled causets. But the labelling is obviously in part %%@
``gauge'', i.e. lacking physical significance. (I say `in part', because of facts like our %%@
example above: $x \prec y \Rightarrow l(x) < l(y)$.) Thus think of adding to the singleton %%@
set $\{ x \}$: either \\
\indent (i): at stage 2, a point $y$ with $x \prec y$; then at stage 3, a point $z$ spacelike %%@
to both $x$ and $y$; or \\
\indent (ii): at stage 2, a point $z$ spacelike to $x$; then at stage 3, a point $y$ with $x %%@
\prec y$ but $z$ spacelike to  $y$ (i.e. {\em not}: $z \prec y$).\\
\indent Intuitively, the contrast between (i) and (ii) is without physical significance, in %%@
the same sort of way that the choice of coordinates in general relativity is without physical %%@
significance (`general covariance' or `diffeomorphism invariance').

The causet approach endorses this intuition, and accordingly requires this sort of difference %%@
to ``cancel out'' in any proposed dynamics.  So the idea is that the probability of any {\em %%@
un}labelled causet $C$ is independent of any attributed ``order of birth''. This is made %%@
precise by imposing an assumption called `discrete general covariance' (DGC). It says that if %%@
$\gamma$ is a path through $\Omega({\cal N})$ from the singleton causet to an $n$-element %%@
causet  $C$ (so $C$ is in rank $n$ of $\Omega({\cal N})$), the product of the transition %%@
probabilities along the links of $\gamma$ is the same as for any other such path.

\subsubsection{Deducing the dynamics}\label{causcondn}
It is a remarkable theorem (due to Rideout and Sorkin (2000)) that DGC, together with an %%@
assumption of relativistic causality, constrains the stochastic dynamics very severely. %%@
Namely:\\
\indent (1): At any stage $n+1$, the probability, $q_n$ say, to add a completely disconnected %%@
element can depend only on $n$, i.e. on the cardinality $n$ of the ``current'' causal set or %%@
the ``current'' rank in $\Omega({\cal N})$. And: \\
\indent (2): The dynamics---the set of all transition probabilities---is given explicitly by %%@
a formula in terms of the $q_n$.

I shall report this theorem, both for its intrinsic interest, and in order to state its  %%@
relativistic causality assumption: that will set the stage for our final return to SEL, in %%@
Section \ref{SELfalse}.

Rideout and Sorkin impose what they call `Bell causality'. It says that the ratio of %%@
probabilities of two transitions depends only on the two corresponding precursor sets. %%@
(Recall that a precursor set is the set of ``ancestors'' of the newborn element.) That is: %%@
Let $C \rightarrow C_1, C \rightarrow C_2$ be two transitions from $C \in {\Omega(n)}$ to %%@
$C_i \in {\Omega(n+1)}$. Then
\begin{equation}
\frac{prob(C \rightarrow C_1)}{prob(C \rightarrow C_2)} =
\frac{prob(B \rightarrow B_1)}{prob(B \rightarrow B_2)}
\label{BellRS99}
\end{equation}
where\\
 (i) $B \in {\Omega(m)}, m \leq n$ is the union of the precursor sets for $C \rightarrow C_1$ %%@
and $C \rightarrow C_2$, and \\
 (ii) $B_i \in {\Omega(m+1)}$ is $B$ with an element added in the manner of the transition $C %%@
\rightarrow C_i$.

Rideout and Sorkin then show that DGC and Bell Causality imply the following two results, (1) %%@
and (2):---\\
\indent (1): The probability to add a completely disconnected element can depend only on the %%@
cardinality $n$ of the ``current'' causet, i.e. the rank in $\Omega({\cal N})$.

To state result (2), we write this probability as $q_n$, and the binomial coefficient %%@
``choose $k$ from $n$'' as $C^n_k$. We then define the parameters
\begin{center}
$t_n := \Sigma^n_{k=0}\; (-)^{n-k} \; C^n_k \; \frac{1}{q_k}$.
\end{center}
Then we have:\\
\indent (2): Setting aside zero probabilities, an arbitrary transition probability, %%@
$\alpha_n$, from ${\Omega(n)}$ to ${\Omega(n+1)}$, with \\
\indent (i): $\varpi$ := the cardinality of the transition's precursor set $S$\\
\indent (ii): $m$ := number of maximal elements in  $S$ (= number of ``parents'' of the new %%@
element)\\
is given by the formula:
\begin{equation}
\alpha_n = \frac{\Sigma^{\varpi}_{l=m} \; C^{\varpi - m}_{\varpi - l} \; t_l}
{\Sigma^n_{j=0} \; C^n_j \; t_j} \;\; .
\label{transprobfixed}
\end{equation}

\subsubsection{The fate of SEL}\label{SELfalse}
Rideout and Sorkin's Bell causality, eq. \ref{BellRS99}, seems intuitively true. For its idea %%@
is that the prospects for a possible transition being realized should not depend on facts (in %%@
this framework, without matter: facts about the causal structure of spacetime) at spacelike %%@
separation from the transition's precursor set. But one naturally asks, especially after %%@
seeing formulations of SEL: surely it is weaker than it needs to be? Why make a statement of %%@
equality of ratios (of transition probabilities), instead of a stronger statement of equality %%@
of probabilities themselves---as SEL does?

That is, one naturally suggests imposing a simpler assumption, about just one transition: %%@
viz. the obvious transcription to the causet approach of the idea of SEL. Thus let us instead %%@
require that for any transition,  $C \rightarrow C_1$:
\begin{equation}
prob(C \rightarrow C_1) =
prob(B \rightarrow B_1)
\label{BellRS99strong}
\end{equation}
where\\
 (i) $B \in {\Omega(m)}, m \leq n$ is the precursor set for $C \rightarrow C_1$, and\\
 (ii) $B_1 \in {\Omega(m+1)}$ is $B$ with an element added in the manner of the transition $C %%@
\rightarrow C_1$.

But Rideout and Sorkin  (and subsequent causet authors) have good reason to impose only their %%@
weaker Bell causality, eq. \ref{BellRS99}. For eq. \ref{BellRS99strong} has the following %%@
defect (Dowker, private communication).  Fix a causet $B$ and consider all the possible %%@
transitions from $B$: $B \rightarrow B_i, i = 1,2, ...., k$. Then of course
\begin{equation}
\Sigma^k_{i = 1} \; prob(B \rightarrow B_i) \; = \; 1.
\label{sumBi}
\end{equation}
Now consider any ``extending'' $C$, with (maybe  many) points spacelike to $B$; (to be %%@
precise, consider any causet $C$ containing a copy of $B$ as a sub-causet, yet with no %%@
elements of $C$ to the future or past of any of the copy of $B$). For any such $C$, there are %%@
corresponding transitions $C \rightarrow C_i$, and so eq. \ref{BellRS99strong}  would demand:
\begin{equation}
\Sigma^k_{i = 1} \; prob(C \rightarrow C_i) \; = \; 1.
\label{sumCi}
\end{equation}
But eq. \ref{sumCi} is unfair to the Elsewhere (within $C$) of $B$! For
$C$ is not obliged to ``grow'' at the current stage from its subset $B$ that we happened to %%@
consider for our instance, eq. \ref{sumBi}, of the law of total probability!

Besides, this defect ``ramifies''. For it is straightforward to show (Dowker, ibid.) that eq. %%@
\ref{BellRS99strong} trivializes the dynamics in that it implies that the only  possible %%@
dynamics is either:\\
\indent (i) with probability 1, only the infinite chain grows; (here the `infinite chain' is %%@
the linearly ordered poset ${\cal N} : = 1 \prec 2 \prec 3 \prec ...$ ) or \\
\indent (ii) with probability 1, only the infinite anti-chain grows; (an anti-chain is the %%@
``completely flat'' poset, with no two elements related by $\prec$).

To end: this discussion prompts a philosopher's puzzle and a technical challenge. The puzzle %%@
(exercise for the reader!) is to say exactly how the contrast between stochastic events in a %%@
fixed spacetime, vs. stochastic events of spacetime structure, makes SEL in the former %%@
framework (as in Sections \ref{sel} to \ref{causeaqft}) immune to the argument just given, %%@
that convicted eq. \ref{BellRS99strong} of being ``unfair to the Elsewhere''.

The technical challenge, suggested by Dowker (and which I leave as a {\em research project} %%@
for the reader) is to formulate, and investigate, other plausible conditions of relativistic %%@
causality for causets. Might one of these deserve the name `SEL'? Work for the future!

\vspace{2 truecm}

Acknowledgements:--- I am grateful to audiences at the Universities of Cambridge, Notre Dame, %%@
Konstanz and Oxford; to Brandon Fogel, Gerd Grasshoff, G\'{a}bor Hofer-Szab\'{o}, Dennis %%@
Lehmkuhl, Thomas M\"{u}ller, Tomasz Placek, Samuel Portmann, Mikl\'{o}s R\'{e}dei, Rafael %%@
Sorkin, Adrian W\"{u}thrich for comments and correspondence (I only wish I could have acted %%@
on all suggestions!); to two referees; and to Adrian Kent and Fay Dowker, respectively, for %%@
teaching me the contents of Sections \ref{kent} and \ref{causets}; and especially to Dennis %%@
Lehmkuhl for the diagrams.

\section{References}\label{refs}
Aspect, A. et al. (1982), `Experimental Tests of Bell inequalities using Time-Varying %%@
Analyzers', {\em Physical Review Letters}, {\bf 49}, pp. 1804-1807.

Bacciagaluppi, G. (2002), `Remarks on Spacetime and Locality in Everett's Interpretation', in %%@
Placek and Butterfield eds. (2002); pp. 105-122.

Bell, J. (1975), The Theory of Local Beables', in Bell (2004); page references are to %%@
reprint.

Bell, J. (2004) {\em Speakable and Unspeakable in Quantum Mechanics}, Cambridge University %%@
Press; second edition; page references are to this edition whose  pagination of the papers %%@
cited matches the first (1987) edition.

Belnap, N. (2005), `A Theory of Causation: {\em Causae causantes} (Originating causes) as %%@
Inus Conditions in Branching Spacetimes', {\em British Journal for the Philosophy of Science} %%@
{\bf 56}, pp. 221-253.  

Belnap, N. and Szab\'{o}, L. (1996), `Branching Spacetime Analysis of the GHZ Theorem', {\em %%@
Foundations of Physics} {\bf 26}, pp. 989-1002.

Berkovitz, J. (1998), `Aspects of Quantum Non-locality I:',  {\em Studies in History and %%@
Philosophy of Modern Physics} {\bf 29B}, pp. 183-222.

Berkovitz, J. (1998a),  `Aspects of Quantum Non-locality II:',  {\em Studies in History and %%@
Philosophy of Modern Physics} {\bf 29B}, pp. 509-546.

Berkovitz. J. (2002), `On Causal Loops in the Quantum Realm', in Placek and Butterfield eds. %%@
(2002); pp. 235-257.

Berkovitz. J. (2007), `Action at a Distance in Quantum Mechanics', {\em Stanford Encyclopedia %%@
of Philosophy}. Available at: http://www.seop.leeds.ac.uk/entries/qm-action-distance/

Brightwell, G., Dowker, H., Garcia, R., Henson, J., and Sorkin, R. (2002), `General %%@
Covariance and the ``Problem of Time'' in a Discrete Cosmology', in K. Bowden (ed.), {\em %%@
Correlations: Proceedings of the ANPA 23 Conference, August 2001, Cambridge, England}, pp. %%@
1-17. Available at: arXiv:gr-qc/0202097

Brightwell, G., Dowker, H., Garcia, R., Henson, J., and Sorkin, R. (2003), ``Observables' in %%@
Causal Set Cosmology', {\em Physical Review} {\bf D67}, 084031. Available at: %%@
arXiv:gr-qc/0210061

Bub, J. (1997), {\em Interpreting the Quantum World}, Cambridge: University Press.

Butterfield, J. (1989), `A spacetime approach to the Bell Inequality', in {\em Philosophical %%@
Consequences of Quantum Theory}, eds. J. Cushing and E. McMullin, Notre Dame University %%@
Press, pp. 114-144.

Butterfield, J. (1992), `Bell's Theorem: what it Takes', {\em British Journal for the %%@
Philosophy of Science} {\bf 42}, pp. 41-83.

Butterfield, J. et al. (1993), `Parameter Dependence  in Dynamical Models of Statevector %%@
Reduction', {\em International Journal of Theoretical Physics} {\bf 32}, pp. 2287-2304.

Butterfield, J. (1994), `Outcome Dependence and Stochastic Einstein Nonlocality' in {\em %%@
Logic and Philosophy of Science in Uppsala}, eds. D Prawitz and D. Westestahl pp. 385-424.

Butterfield, J. (1996) `Vacuum Correlations and Outcome Dependence in
Algebraic Quantum Field Theory, in {\em Fundamental Aspects of Quantum
Theory}, eds. D. Greenberger and A. Zeilinger, New York Academy of
Sciences, New York, pp. 768-785.

Butterfield, J. (2007), `Reconsidering Relativistic Causality', forthcoming in {\em %%@
International Studies in the Philosophy of Science}.

Cramer, J. (1986), `The Transactional Interpretation of Quantum Mechanics', {\em Reviews of %%@
Modern Physics} {\bf 58}, pp. 647-687.

Dowker, H. F. (2005), `Causal Sets and the Deep Structure of Spacetime',  Available at: %%@
arXiv:gr-qc/0508109

Earman, J. (2006), `Pruning some Branches from `Branching Spacetimes', preprint.

Fahmi, A. and Golshani, M. (2006), `Locality and the Greenberger-Horne-Zeilinger theorem', %%@
Available at: arXiv:quant-ph/0608049

Geroch, R. and Horowitz, G. (1979), `Global structure of spacetimes', in {\em General %%@
Relativity: an Einstein Centennial Survey}, ed. S. Hawking and W. Israel, Cambridge %%@
University Press,  pp. 212-293.

Grasshoff, G. et al. (2005), `Minimal Assumption Derivation of a Bell-type Inequality', {\em %%@
British Journal for the Philosophy of Science} {\bf 56}, pp. 663-680.

Halvorson, H. and Clifton, R. (2000), `Generic Bell correlation between arbitrary local %%@
algebras in quantum field theory', {\em Journal of Mathematical Physics} {\bf 41}, pp. %%@
1711-1717; reprinted in R. Clfton, {\em Quantum Entanglements}, (eds. J. Butterfield and H. %%@
Halvorson), Oxford University Press, 2004. Available at: arXiv:math-ph/9909013

Hawking, S. and Ellis, G. (1973), {\em The Large-Scale Structure of Spacetime}  Cambridge: %%@
University Press.

Hellman, G. (1982), `Stochastic Einstein Locality and the Bell Theorems', {\em Synthese} {\bf %%@
58}, pp. 461-504.

Henson, J. (2005), `Comparing Causality Principles', {\em Studies in History and Philosophy %%@
of Modern Physics} {\bf 36B}, pp. 519-543.

Hofer-Szab\'{o}, G.  et al. (1999), `On Reichenbach's common cause principle and %%@
Reichenbach's notion of common cause', {\em British Journal for the Philosophy of Science} %%@
{\bf 50}, pp. 377-399. Available at; arXiv:quant-ph/9805066

Hofer-Szab\'{o}, G.  et al. (2002), `Common-causes are not common common-causes', {\em %%@
Philosophy of Science} {\bf 69}, pp. 623-633. Available at; %%@
http://philsci-archive.pitt.edu/archive/00000353/

Hofer-Szab\'{o}, G. (2007), `Separate vs. {\em common}-common-cause-type derivations of the %%@
Bell inequalities', forthcoming in {\em Synthese}. 

Kent, A. (2005), `Causal Quantum Theory and the Collapse Locality Loophole', {\em Physical %%@
Review A} {\bf 72}, 012107. Available at: arXiv:quant-ph/0204104

Kent, A. (2005a), `A Proposed Test of the Local Causality of Spacetime'.
 Available at: arXiv:gr-qc/0507045

Kowalski, T. and Placek, T. (1999) `Outcomes in Branching Spacetime and GHZ-Bell Theorems' %%@
{\em British Journal for the Philosophy of Science} {\bf 50}, pp. 349-375.

Lewis, D. (1980) `A Subjectivist's Guide to Objective Chance', in R.C. Jeffrey ed. {\em %%@
Studies in Inductive Logic and Probability} volume II University of California Press; %%@
reprinted in Lewis' {\em Philosophical Papers}, volume II, Oxford University Press 1986; page %%@
references to reprint.

Lewis, D. (1986), {\em On the Plurality of Worlds}, Oxford: Blackwell.

Loeve, M. (1963), {\em Probability Theory}, Princeton; van Nostrand.

M\"{u}ller, T. (2005), `Probability Theory and Causation: a Branching Spacetimes Analysis', %%@
{\em British Journal for the Philosophy of Science} {\bf 56}, pp. 487-520.  

Muller, F. and Butterfield, J. (1994), `Is Algebraic Lorentz-covariant Quantum Field Theory %%@
Stochastic Einstein Local?', {\em Philosophy of Science} {\bf 61}, pp. 457-474.

Norton, J. (2003),`Causation as Folk Science', {\em Philosophers' Imprint} {\bf 3} \\
 http://www.philosophersimprint.org/003004/; to be reprinted in H. Price and R. Corry, {\em %%@
Causation and the Constitution of Reality}, Oxford: University Press.

Norton, J. (2006), `Do the Causal Principles of Modern Physics Contradict Causal %%@
Anti-fundamentalism?', to appear in {\em Causality: Historical and Contemporary},
eds. P. K. Machamer and G. Wolters, Pittsburgh: University of Pittsburgh Press. Available at: %%@
http://philsci-archive.pitt.edu/archive/00002735/

Placek, T. (2000), `Stochastic Outcomes in Branching Spacetime: Analysis of Bell's Theorem', %%@
{\em British Journal for the Philosophy of Science} {\bf 51} , pp. 445-475.

Placek, T. (2000a), {\em Is Nature Deterministic?}, Jagiellonian University Press.

Placek, T. (2002), `Partial Indeterminism is Enough: a branching analysis of Bell-type %%@
inequalities', in Placek and Butterfield eds. (2002); pp. 317-342.

Placek, T. (2004), `Screening-off Conditions in Bell's Theorem: a Branching Spacetimes %%@
Analysis', in L. Bihounek and M. Bilkova eds., {\em Logica Yearbook 2004}, Prague: Filosofia, %%@
pp. 243-354.

Placek, T. and Butterfield, J., eds. (2002), {\em Non-locality and Modality}, Dordrecht: %%@
Kluwer Academic (Nato Science Series, vol. 64).

Portmann, S. and W\"{u}thrich, A. (2007), `Minimal assumption derivation of a weak %%@
Clauser-Hourne inequality', forthcoming in {\em Studies in History and Philosophy of Modern %%@
Physics}. Available at: arXiv:quant-ph/0604216

Price, H. (1996), {\em Time's Arrow and Archimedes' Point}, Oxford: University Press.

R\'{e}dei, M. (1991), `Bell's Inequalities, Relativistic Quantum Field Theory and the Problem %%@
of Hidden Variables', {\em Philosophy of Science} {\bf 58}, pp. 628-638.

R\'{e}dei, M. (2002), `Reichenbach's common cause principle and quantum correlations', in %%@
Placek and Butterfield eds. (2002); pp. 259-270.

R\'{e}dei, M. and Summers, S. (2002) `Local Primitive Causality and the Common Cause %%@
Principle in Quantum Field Theory', {\em Foundations of Physics} {\bf 32} pp. 335-355. %%@
Available at: arXiv:quant-ph/0108023

Redhead, M. (1987), {\em Incompletness, Nonlocality and Realism}, Oxford: University Press.

Redhead, M. (1995), `More Ado about Nothing', {\em Foundations of Physics}, {\bf 25}, pp. %%@
123-137.

Reichenbach, H. (1956), {\em The Direction of Time}, Berkeley: University of California %%@
Press.

Rideout, D. and Sorkin, R. (2000), `A Classical Sequential Growth Dynamics for Causal Sets', %%@
{\em Physical Review} {\bf D61}, 024002. Available at: arXiv:gr-qc/9904062

Santos, E. (2005), `Bell's Theorem and the Experiments: Increasing empirical support for %%@
local realism?', {\em Studies in the History and Philosophy of Modern Physics} {\bf 36B}, pp. %%@
544-565.

Seevinck, M. (forthcoming), `Deriving standard Bell inequalities from non-locality and its %%@
repercussions for the (im)possibility of doing experimental metaphysics'.

Shimony, A. (2004), `Bell's Theorem', {\em Stanford Encyclopedia of Philosophy}. Available %%@
at: http://www.seop.leeds.ac.uk/entries/bell-theorem/

Socolovsky, M. (2003), `Bell inequality, non-locality and analyticity', {\em Physics Letters %%@
A}, {\bf 316}, pp. 10-16. Available at: arXiv:quant-ph/0305135 

Su\'{a}rez, M. (2007), `Causal Inference in Quantum Mechanics: A Reassessment',
in F. Russo and J. Williamson (eds.), {\em Cauaality and Probability in
the Sciences}, London College Texts, pp. 65-106.

Summers, S. (1990), `On the Independence of Local Algebras in Quantum Field Theory', {\em %%@
Reviews in Mathematical Physics} {\bf 2}, pp. 201-247.

Szab\'{o}, L. (2000), `On an attempt to resolve the EPR-Bell paradox via Reichenbachian %%@
concept of common cause', , {\em International Journal of Theoretical Physics} {\bf 39}, pp. %%@
901-911. Available at: arXiv: quant-ph/9806074

Timpson, C. and Brown, H. (2002), 'Entanglement and Relativity' in {\em Understanding %%@
Physical Knowledge},  R. Lupacchini and V. Fano (eds.), 2002; University of Bologna, CIUEB. %%@
Available at: arXiv:quant-ph/0212140, and at:\\ %%@
http://philsci-archive.pitt.edu/archive/00001618/

Uffink, J. (1999), `The Principle of the Common Cause faces the Bernstein Paradox', {\em %%@
Philosophy of Science} {\bf 66}, pp. S512-S525 (Supplement: Proceedings of 1998 Conference).

Wald, R. (1984), {\em General Relativity}, Chicago: University of Chicago Press.

Weihs, G. et al. (1998) `Violation of Bell inequality under strict Einstein locality %%@
conditions', {\em Physical Review Letters} {\bf 81}, pp. 5039-5043.

\end{document}